\newcommand{\symImpedance}{z}
\newcommand{\symPhaseA}{{a}}
\newcommand{\symPhaseB}{{b}}
\newcommand{\symPhaseC}{{c}}
\newcommand{\symPhaseN}{{n}}
\newcommand{\seriesss}{\text{{s}}}
\newcommand{\indexGridLines}{{l}}
\newcommand{\zijksAA}[0]{{\symImpedance_{ \indexGridLines , \symPhaseA \symPhaseA}^{\seriesss}    }}
\newcommand{\zijksAB}[0]{{\symImpedance_{ \indexGridLines , \symPhaseA \symPhaseB}^{\seriesss}    }}
\newcommand{\zijksAC}[0]{{\symImpedance_{ \indexGridLines , \symPhaseA \symPhaseC}^{\seriesss}    }}
\newcommand{\zijksBA}[0]{{\symImpedance_{ \indexGridLines , \symPhaseB \symPhaseA}^{\seriesss}    }}
\newcommand{\zijksBB}[0]{{\symImpedance_{ \indexGridLines , \symPhaseB \symPhaseB}^{\seriesss}    }}
\newcommand{\zijksBC}[0]{{\symImpedance_{ \indexGridLines , \symPhaseB \symPhaseC}^{\seriesss}    }}
\newcommand{\zijksCA}[0]{{\symImpedance_{ \indexGridLines , \symPhaseC \symPhaseA}^{\seriesss}    }}
\newcommand{\zijksCB}[0]{{\symImpedance_{ \indexGridLines , \symPhaseC \symPhaseB}^{\seriesss}    }}
\newcommand{\zijksCC}[0]{{\symImpedance_{ \indexGridLines , \symPhaseC \symPhaseC}^{\seriesss}    }}

\newcommand{\zijksAN}[0]{{\symImpedance_{ \indexGridLines , \symPhaseA \symPhaseN}^{\seriesss}    }}
\newcommand{\zijksBN}[0]{{\symImpedance_{ \indexGridLines , \symPhaseB \symPhaseN}^{\seriesss}    }}
\newcommand{\zijksCN}[0]{{\symImpedance_{ \indexGridLines , \symPhaseC \symPhaseN}^{\seriesss}    }}
\newcommand{\zijksNA}[0]{{\symImpedance_{ \indexGridLines , \symPhaseN \symPhaseA }^{\seriesss}    }}
\newcommand{\zijksNB}[0]{{\symImpedance_{ \indexGridLines ,\symPhaseN \symPhaseB }^{\seriesss}    }}
\newcommand{\zijksNC}[0]{{\symImpedance_{ \indexGridLines , \symPhaseN\symPhaseC }^{\seriesss}    }}

\newcommand{\zijksNN}[0]{{\symImpedance_{ \indexGridLines , \symPhaseN \symPhaseN}^{\seriesss}    }}

\documentclass[journal,12pt,onecolumn,draftclsnofoot,]{IEEEtran}

\usepackage[cmex10]{amsmath}
\usepackage[utf8]{inputenc}
\usepackage{multirow}
\usepackage{soul}
\usepackage{algorithm}
\usepackage{algpseudocode}
\usepackage{graphicx}
\usepackage{dsfont}
\usepackage{xcolor}
\usepackage{cite}

\usepackage[utf8]{inputenc}
\usepackage{mathtools}
\usepackage[mathscr]{euscript}
\usepackage{amsmath}
\usepackage{amssymb}
\usepackage{amsfonts}
\usepackage{amssymb}
\usepackage{amsmath}
\usepackage{verbatim}
\usepackage[T1]{fontenc}
\usepackage[utf8]{inputenc}
\usepackage{latexsym}
\usepackage{enumerate}
\usepackage{indentfirst}
\usepackage{calligra}
\usepackage{ulem}
\usepackage{graphicx}
\usepackage{array}
\usepackage{float}
\usepackage{bbm}

\usepackage{mleftright}

\makeatletter
\let\old@ps@headings\ps@headings
\let\old@ps@IEEEtitlepagestyle\ps@IEEEtitlepagestyle
\def\psccfooter#1{%
    \def\ps@headings{%
        \old@ps@headings%
        \def\@oddfoot{\strut\hfill#1\hfill\strut}%
        \def\@evenfoot{\strut\hfill#1\hfill\strut}%
    }%
    \def\ps@IEEEtitlepagestyle{%
        \old@ps@IEEEtitlepagestyle%
        \def\@oddfoot{\strut\hfill#1\hfill\strut}%
        \def\@evenfoot{\strut\hfill#1\hfill\strut}%
    }%
    \ps@headings%
}
\makeatother

\usepackage{longtable}
\usepackage{soul}

\usepackage{multirow}
\usepackage{soul,color}
\usepackage{graphicx}
\usepackage{dsfont}
\usepackage{subcaption}

\usepackage[utf8]{inputenc}
\usepackage{mathtools}
\usepackage[mathscr]{euscript}
\usepackage{amsmath}
\usepackage{amssymb}
\usepackage{amsfonts}
\usepackage{verbatim}
\usepackage[T1]{fontenc}
\usepackage[utf8]{inputenc}
\usepackage{latexsym}
\usepackage{enumerate}
\usepackage{indentfirst}
\usepackage{calligra}
\usepackage{ulem}

\usepackage{array}
\usepackage{float}
\usepackage{bbm}

\usepackage{geometry}
\usepackage{eurosym}

\usepackage{hyperref}

\usepackage{caption}

\usepackage{tikz}
\usepackage{pgfplots}

\pgfplotsset{compat=1.8}
\usepgfplotslibrary{statistics}
\pgfmathdeclarefunction{fpumod}{2}{%
        \pgfmathfloatdivide{#1}{#2}%
        \pgfmathfloatint{\pgfmathresult}%
        \pgfmathfloatmultiply{\pgfmathresult}{#2}%
        \pgfmathfloatsubtract{#1}{\pgfmathresult}%
        \pgfmathfloatifapproxequalrel{\pgfmathresult}{#2}{\def\pgfmathresult{5}}{}%
    }

\usepackage{tabularx}
\usepackage{flushend}
\usepackage{textcomp}
\usepackage{xcolor}
\usepackage{import}

\usetikzlibrary{trees,decorations,shadows}
\tikzset{level 1/.style={sibling angle=45,level distance=4mm}}
\usetikzlibrary{arrows.meta}
\usepackage{forest}
\usetikzlibrary{external}

\let\oldtikzexternalgetnextfilename\tikzexternalgetnextfilename \renewcommand{\tikzexternalgetnextfilename}[1]{\oldtikzexternalgetnextfilename{#1}\expandafter\tikzsetnextfilename\expandafter{#1}}

\usepackage{pgfplotstable}
\usepackage[outline]{contour}
\contourlength{1.2pt}

\pgfplotsset{compat=1.13} 

\usetikzlibrary{spy}
\usetikzlibrary{calc}
\usetikzlibrary{fadings}
\usetikzlibrary{patterns}
\usetikzlibrary{shadows}
\usetikzlibrary{mindmap}
\usetikzlibrary{backgrounds}
\usetikzlibrary{shapes.symbols}
\usetikzlibrary{shapes.multipart}
\usetikzlibrary{shapes.geometric}
\usetikzlibrary{automata,positioning}
\usetikzlibrary{decorations.fractals} 
\usetikzlibrary{decorations.markings}
\usetikzlibrary{decorations.pathreplacing}
\usetikzlibrary{decorations.pathmorphing}
    
\usepackage{bm}

\usepackage{nomencl}
\makenomenclature
\tikzset{edge from parent/.style={segment angle=10,draw}}

\tikzset{
  my rounded corners/.append style={rounded corners=2pt},
}

\def\BibTeX{{\rm B\kern-.05em{\sc i\kern-.025em b}\kern-.08em
    T\kern-.1667em\lower.7ex\hbox{E}\kern-.125emX}}

\renewcommand{\nomgroup}[1]{%
     \ifthenelse{\equal{#1}{O}}{\item[\textit{Operators}]}{%
        \ifthenelse{\equal{#1}{I}}{\item[\textit{Indices}]}{%
            \ifthenelse{\equal{#1}{A}}{\item[\textit{Acronyms}]}{%
            `\ifthenelse{\equal{#1}{V}}{\item[\textit{Variables and parameters}]}{}}}}}
\usepackage{scalerel}
\usetikzlibrary{svg.path}
\definecolor{orcidlogocol}{HTML}{A6CE39}
\tikzset{
    orcidlogo/.pic={
        \fill[orcidlogocol] svg{M256,128c0,70.7-57.3,128-128,128C57.3,256,0,198.7,0,128C0,57.3,57.3,0,128,0C198.7,0,256,57.3,256,128z};
        \fill[white] svg{M86.3,186.2H70.9V79.1h15.4v48.4V186.2z}
        svg{M108.9,79.1h41.6c39.6,0,57,28.3,57,53.6c0,27.5-21.5,53.6-56.8,53.6h-41.8V79.1z M124.3,172.4h24.5c34.9,0,42.9-26.5,42.9-39.7c0-21.5-13.7-39.7-43.7-39.7h-23.7V172.4z}
        svg{M88.7,56.8c0,5.5-4.5,10.1-10.1,10.1c-5.6,0-10.1-4.6-10.1-10.1c0-5.6,4.5-10.1,10.1-10.1C84.2,46.7,88.7,51.3,88.7,56.8z};
    }
}

\newcommand\orcidicon[1]{\href{https://orcid.org/#1}{\mbox{\scalerel*{ \begin{tikzpicture}[yscale=-1,transform shape]
                \pic{orcidlogo};
                \end{tikzpicture}
            }{|}}}}

\begin{document}

\title{Consensus based phase connectivity identification for distribution network with limited observability
}


\author{Md~Umar~Hashmi$^{1}$~\orcidicon{0000-0002-0193-6703},~David Brummund$^{2}$,~Rickard Lundholm$^{1}$,  Arpan~Koirala$^{1}$~\orcidicon{0000-0003-4826-7137}, and~Dirk~Van~Hertem$^{1}$~\orcidicon{0000-0001-5461-8891}
\thanks{Corresponding author email: mdumar.hashmi@kuleuven.be}
\thanks{$^{1}$Md~Umar~Hashmi, Rickard Lundholm, Arpan~Koirala and Dirk~Van~Hertem are with KU Leuven, division Electa \& EnergyVille, Genk, Belgium}
\thanks{$^{2}$David Brummund is with MITNETZ STROM, Germany }
} 

\maketitle

\begin{abstract}
The mitigation of distribution network (DN) unbalance and the use of single-phase flexibility for congestion mitigation requires accurate phase {connection} information, which is often not available.
For a large DN, the na\"{i}ve phase identification proposed in the majority of the prior works using a single {voltage} reference does not scale well {for a multi-feeder DN}.
We present a consensus algorithm-based phase {identification} mechanism which uses multiple {three-phase reference points} to improve the prediction of phases.
Due to the absence of real measurements for {a real-suburban German DN}, the algorithms are developed and evaluated over synthetic data using a digital twin. 
To utilize strongly correlated measurements, the DN is clustered into zones. 
{We observe those reference measurements located in the same zone as the single-phase consumer leads to
accurate prediction of DN phases.}
Four consensus algorithms are developed and compared. Using numerical results, we recommend the most robust phase identification mechanism.
In our evaluation, measurement error, and the impact of the neutral conductor are also assessed.
We assume limited DN observability and apply our findings to a German DN without smart meters, but only less than 8\% of nodes have measurement boxes along with single-phase consumers with a home energy management system. 
Voltage time series for 1 month (hourly sampled) is utilized.
The numerical results indicate that for 1\% accuracy class measurement, the phase connectivity of 308 out of 313 single-phase consumers in a German DN can be identified.
Further, we also propose metrics quantifying the goodness of the phase identification. The phase identification framework based on consensus algorithms for DN zones is scalable for large DN and robust towards measurement errors as the estimation is not dependent on a single measurement point.

\end{abstract}


\tableofcontents

\pagebreak


\section{Introduction}
The low voltage distribution network (DN) in Europe consists predominantly of single phase (1-$\phi)$ loads, inverter interfaced PV, storage, and electric vehicle charging infrastructure.
Often the phase connectivity of such resources is not accurately known.
This lack of DN observability will restrict the monitoring and control of DN imbalances.
Existing DN consists of multiple measurements at the feeder level and end of the feeder, which provides utilities with some degree of observability.
The goal of the paper is to develop a scalable and robust phase connectivity identification (PCI) framework that considers multiple measurement points for improving the PCI.

\begin{table*}
\centering
\tiny
\caption{\small{Literature review on non-intrusive phase identification}}
\label{tab:phaselit}
\begin{tabular}{p{6mm}|p{30mm}|p{52mm}|p{42mm}|p{22mm}|p{13mm}} 
\hline
Ref            & Measurement dependency                                                               & Proposed solution                                                                                                                                                                                 & Remarks                                                                                               & Methodology                      & Input      \\ 
\hline
\hline
\cite{pi1:blakely2019spectral}               & AMI voltage time series, partially incorrect phase label information              & Spectral clustering with a sliding window; does not require substation measurement                                                                                                                & 91\% accuracy, Google street view analysed for phase identification                                       & Clustering / Unsupervised ML     & voltage               \\ 
\hline
\cite{pi77:liu2020practical}                 & Voltage magnitude   (denoted as $|V|$)                                                   & Spectral clustering is utilized and MILP model is used for unbalance mitigation.                                                                                                                  & 156 user DN in China is used for validation. Majority rule is applied to over predictions over new data. & Clustering / Unsupervised ML     & voltage               \\ 
\hline
\cite{pi3:ni2017phase}                & Voltage magnitude                                                   & k-means clustering with Gaussian Mixture Model algorithm for phase id                                                                                                                             & 91\% accurate; with salient features accuracy is 100\%                                                & Clustering / Unsupervised ML     & voltage               \\ 
\hline
\cite{pi34:mitra2015voltage}                 & Voltage magnitude                                                     & k-means clustering is used. Use multiple references with \textbf{Majority rule} based estimation.                                                                                                     & 90\% accuracy                                                                                         & Clustering / Unsupervised ML     & voltage               \\ 
\hline 
\cite{pi23:zaragoza2022denoising}                 & Voltage magnitude                                                     & k-medoids clustering is with denoised data                                                                                                      & Singular value decomposition is used for denoising                                                                                         & Clustering / Unsupervised ML     & voltage               \\ 
\hline
\cite{pix1:wang2016phase}               & Voltage magnitude                                                      & 
principal
components are used to extract feature vectors over which
constrained k-means clustering is applied                                                                                                & 90\% accuracy                                                                                          & Clustering / Unsupervised ML     & voltage               \\ 
\hline
\cite{simonovska2021phase}                 & Voltage magnitude                                                     & k-means clustering with principal component analysis                                                                                                      & Phase identification is applied for multiple days separately and a majority rule is applied                                                                                       & Clustering / Unsupervised ML     & voltage               \\ 
\hline
\cite{pi5:xu2016phase}                  & Active power measurements, substation as reference                                & extract distinct features from load profiles and correlate with phase load; limitation: high granularity data needed                                                                              & 93\% accuracy with 10\% SMs in DN; results compared with \cite{pi7:arya2011phase}                                          & Correlation                      & power                 \\ 
\hline
\cite{pi10:pezeshki2012consumer}              & Voltage magnitude                                                     & Correlation based; the salient features of the time series are extracted.                                                                                                                      & Large enough data sheet leads to 100\% accuracy for 75 consumer DN                                     & Correlation                      & voltage               \\ 
\hline
\cite{pi17:olivier2017automatic}                 & Voltage magnitude                                                    & Relies on graph theory and the notion of maximum spanning tree. Correlation based PCI for a {four wire DN}.                                                       & \textbf{Closer the measurement points are geographically, the stronger the correlation between the voltages}   & Correlation                      & voltage               \\ 
\hline
\cite{pi6:vycital2019phase}                  & Voltage magnitude                                                      & Difference matrix is created                                                                                                                                                                      & 82\% accuracy                                                                                         & Correlation                      & voltage               \\ 
\hline
\cite{pi4:olivier2018phase}                  & Voltage magnitude time series                                                     & Correlation between voltage measurements of SMs with constrained k-means                                                                      &           substation voltage is used as reference for phase identification                                                                                            & Correlation with unsupervised ML & voltage               \\ 
\hline
\cite{pi9:hoogsteyn2022low}                 & Active power and voltage time series data                                                            & Correlationship with clustering is performed. {Ensemble learning combines voltage and power-based estimation results. }                                                                                                                                                             &            Impact of different SM accuracy class is evaluated                                                                                           & Correlation with unsupervised ML & voltage and power     \\ 
\hline
\cite{pi7:arya2011phase}                  & Active power time series at the transformer and consumers                             & integer programming along with branch and bound search algorithm. Access sensitivity of the ratio of measurement points  and total number of consumers                                             & MILP based solution depends on the principle of conservation of energy.                               & Optimization                     & power                 \\ 
\hline
\cite{pi32:heidari2021phase}                 & P, Q and {$|V|$} measurements                                                                & MILP with Bender's decomposition is used. Accuracy of phase id is governed by number of data points, SM class, data resolution                                                                    & For large EU feeder, the runtime with 5\% SM error is 39.2 hours. Difficult to scale.         & Optimization                     & power (P \& Q) and voltage                \\ 
\hline
\cite{vanin2022phase}  & P, Q, $|V|$ measurements                                                                & Utilize state-estimation with MILP, also considers errors in layout information                                      & Data needs are smaller compared to statistical and ML-based techniques         & Optimization                     & power and voltage                 \\ 
\hline
\cite{pi18:xiaoqing2018phase}                 & Active power time series                                                          & LASSO based data driven approach. Also considers SM accuracy class                                                                                                                                & 97\% accuracy with 60\% SMs; LASSO immune to noise unlike \cite{pi7:arya2011phase}                                         & Statistical or ML                & power                 \\ 
\hline
\cite{pi14:jayadev2016novel, pi12:wen2015phase, pi2:pappu2017identifying}      & Energy measurement time series                                                    & data-driven approach with Principal component analysis \& graph theory interpretations                                                                                             & Also considers noisy data                                                                             & Statistical or ML                & power                 \\ 
\hline
\cite{pi8:zhou2021consumer}                  & Voltage magnitude time series                                                     & Develops multi-dimensional calibration in phase id based on voltage characteristics in LVDN                                                                                                        & Observe that voltage characteristics are more robust under incomplete data                            & Statistical or ML                & voltage               \\ 
\hline
\cite{pi30:short2012advanced}                 & Voltage magnitude                                                     & Linear regression and voltage drop relationship for phase identification                                                                                                                                                        &  \textbf{Observe close measurement are strongly correlated }                                                                                                    & Statistical or ML                & voltage               \\ 
\hline
\cite{pi12:wen2015phase}                & Voltage phasor time series measured using microPMUs                               & Phase id analyses cross correlations over voltage magnitudes and detects the phase angle difference between reference and test nodes                                                              & multi-phase connections are also considered. Fine resolution of 120 samples per sec is used           & Statistical or ML                & voltage phasor        \\ 
\hline
\cite{pi21:padullaparti2019considerations}                & P, $|V|$ time series                                                             & Using statistical analysis of AMI data over a day for DN with PV. Regression model is used to model substation voltage using nodal P, V and substation power.                                     & Explore data needs, granularity and impact of PV penetration levels                                   & Statistical or ML                & voltage and power     \\ 
\hline
\cite{pi22:foggo2018comprehensive}                & Voltage magnitude and phase info of a small representative set & Train an ML model of constrained function of voltage time series. Since manual measurements are needed thus may not be scalable                                                                    & 5\% selected representative set leads to accuracy of 91.9\%                                           & Statistical or ML                & voltage               \\ 
\hline
\cite{pi20:liao2019unbalanced}                 & Voltage phasor time series                                                        & Use sequence component for phase identification                                                                                                                                                   &          Also utilize SM data for learning the topology of the DN                                                                                             & Statistical or ML                & voltage phasor        \\ 
\hline
\cite{pi13:foggo2019improving}                &    Voltage magnitude                                                                              & Supervised machine learning with theory of information loss                                                                                                                                       & Accuracy up to 97\%                                                                                    & Supervised ML                    &                       \\ 
\hline
\cite{pi101:bariya2021guaranteed}               & Voltage phasor time series                                                        & Topology and phase identification using linearized model of three-phase unbalanced DN.                                          & 120 Hz PMU measurements are used. Data collected for 1 second to 1 minute is used for estimation                                             & Statistical or ML                & voltage phasors       \\ 
\hline
\multicolumn{1}{l}{} & \multicolumn{1}{l}{}                                                             & \multicolumn{1}{l}{}                                                                                                                                                                              & \multicolumn{1}{l}{}                                                                                  & \multicolumn{1}{l}{}             & \multicolumn{1}{l}{} 
\end{tabular}
\end{table*}


\begin{figure}[!htbp]
	\center
	\includegraphics[width=0.98\linewidth]{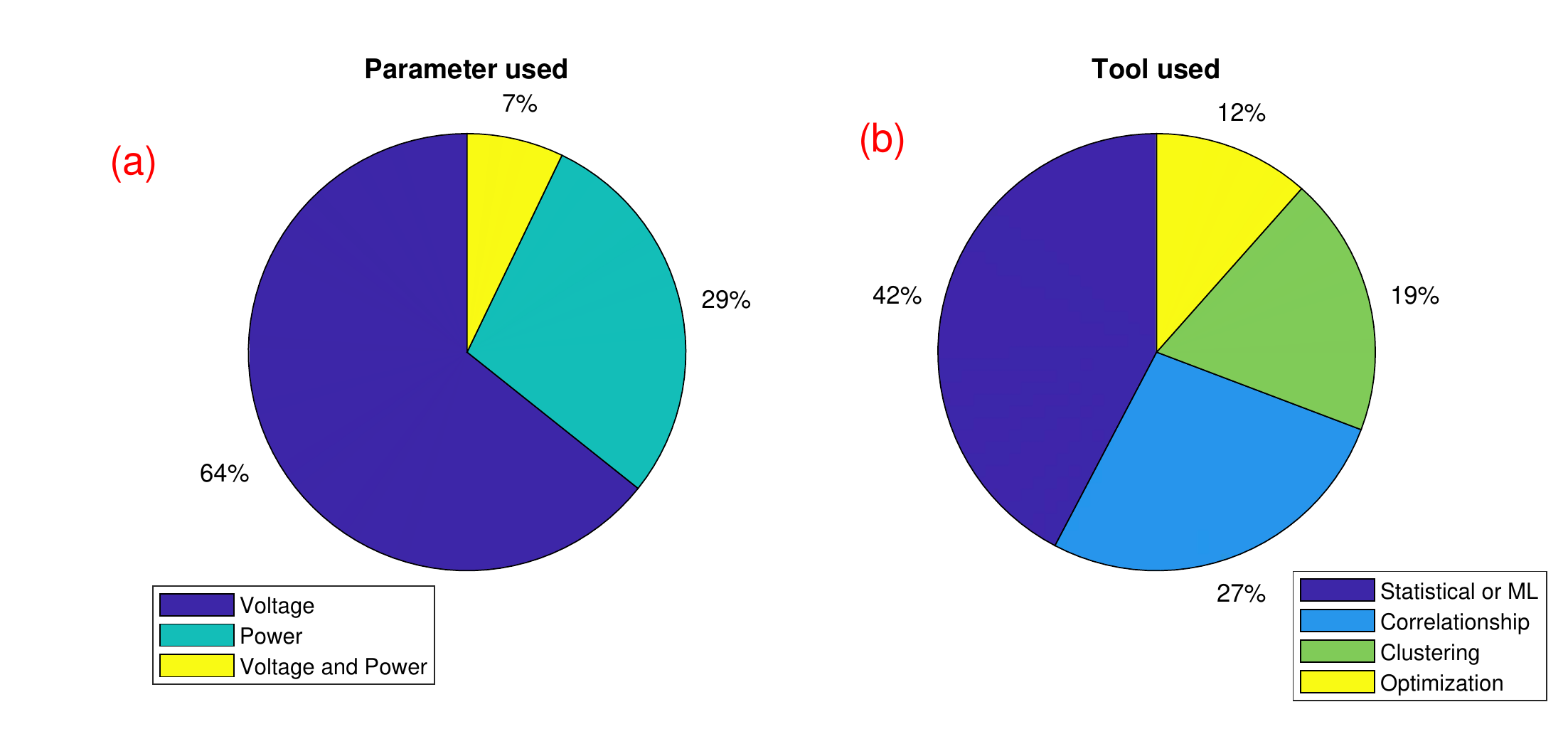}
	\vspace{-7pt}
	\caption{\small{Classifying of phase connectivity identification literature based on parameter(s) and tool(s) used.}}
	\label{fig:literature}
\end{figure}

Phase identification methodologies can be broadly classified as intrusive and non-intrusive. As the name suggests, intrusive methods require manual identification {of phases and are often labor-intensive and/or hardware-based \cite{kolwalkar2014phase}}. On the other hand, non-intrusive methods are often data-driven. A brief summary of non-intrusive phase identification methods is detailed in Table \ref{tab:phaselit}.
These data-driven methods can be classified based on the parameter used and tool used for phase identification.
For the literature summarized in Table \ref{tab:phaselit}, the classification is presented in Fig. \ref{fig:literature}. 
From Fig. \ref{fig:literature}, we observe that 64\% of existing works utilize voltage magnitude time series and approximately 27\% utilize correlation as a tool for PCI. 
In this work, we also utilize these widely used {techniques} for developing a consensus-based phase identification framework {using voltage time series and correlation as the tool. Voltage time series-based phase identification is more robust to limited observability in a DN, also observed in \cite{matijavsevic2022voltage}. The power-based methods rely on the law of conservation of energy and require a high degree of observability in a DN.}

\textit{Motivation}:
With the increasing generation from renewable energy sources and the growing addition of flexible loads like electric vehicles, heat pumps, etc. congestions, voltage violations and phase imbalances in the grid will likely become more frequent. Therefore, suitable corrective measures must be found and implemented. An important basis for mitigation measures is the detection of congestions and thus, firstly, {improving the network model of the} low-voltage grid, which is still largely unmonitored in {many parts of the world} today. 
{Accurate network topology is assumed to be known in many works \cite{wang2020optimal, HASHMI2022108608}}.
However, this assumption is not accurate in the case of \href{https://euniversal.eu/}{EUniversal's} demo networks for the German DN.


\subsection{Observations of this paper}
\begin{figure*}[!htbp]
	\center
	\includegraphics[width=0.78\linewidth]{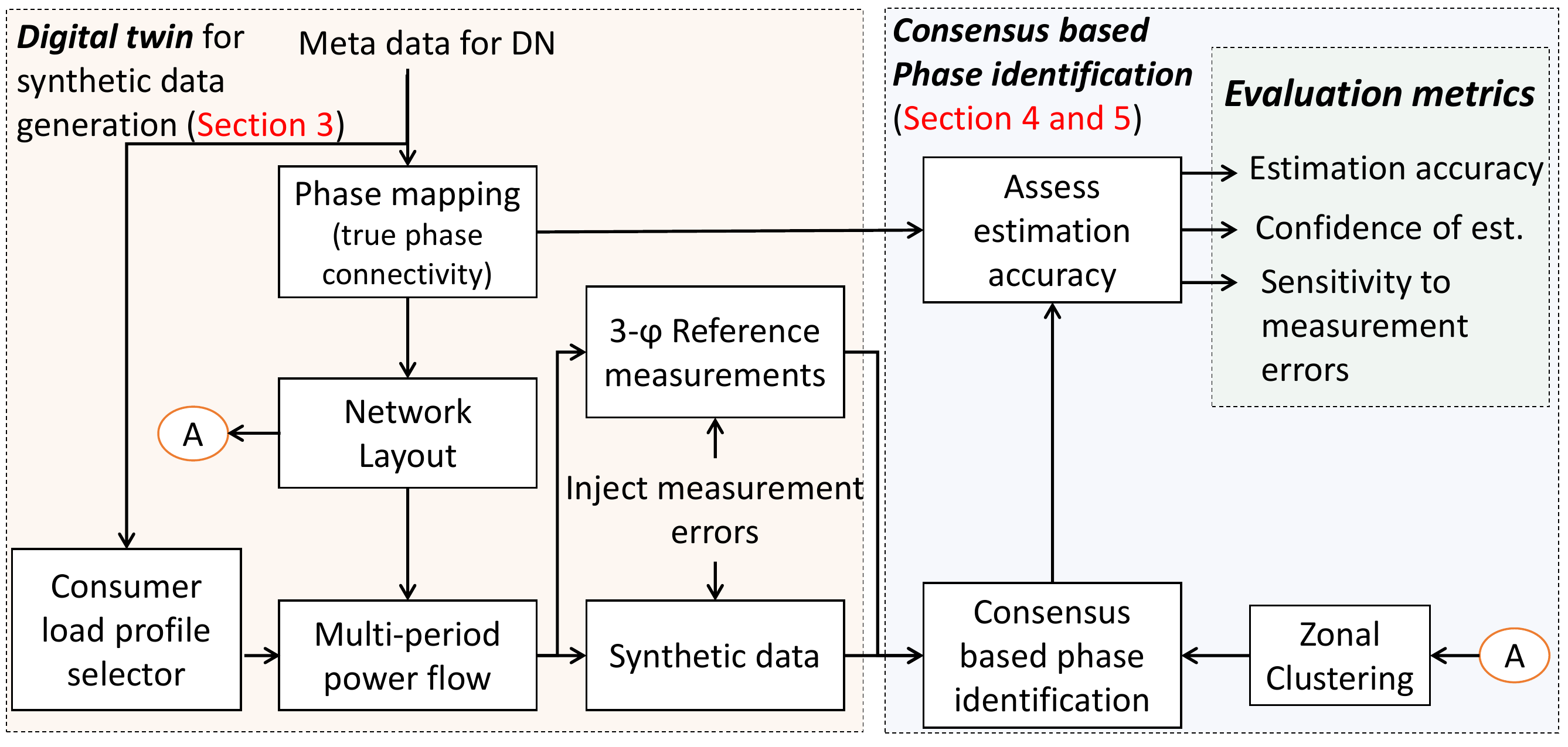}
	\vspace{-2pt}
	\caption{\small{Consensus-based phase identification, synthetic data generation and metrics used}}
	\label{fig:structure}
\end{figure*}

The contributions and observations of the paper are as follows:
\begin{itemize}
\item A tailor-made solution for the German DSO, \href{https://www.mitnetz-strom.de/}{Mitnetz Strom}, is proposed for phase identification which considers multiple measurements in a DN zone for improving the PCI accuracy, this is detailed in Fig. \ref{fig:structure}.
{The proposed framework can also be applied for other DNs with limited DN observability.}
\item {Twelve} phase identification models are benchmarked over the na\"{i}ve model. The na\"{i}ve model considers only one {$3-\phi$ measurement} reference in a DN for phase estimation. The proposed phase identification models build a consensus among multiple measurements for robust phase estimation.
\item Metrics are proposed for evaluating phase identification models using (a1) accuracy of estimation, (a2) confidence factor, and (a3) sensitivity towards measurement errors. Metrics (a2) and (a3) provide a qualitative metric for evaluating estimation accuracy.
\item A detailed description is provided for synthetic data generation, which utilizes 
{an example suburban LV DN grid model of Mitnetz Strom.}
{This is also crucial for DNs with limited or no historical measurement data. }
\item Four case studies are performed in the numerical results. {
The performance of the na\"{i}ve model is used for benchmarking and comparing the proposed phase identification algorithms.}
\begin{itemize}
\item Firstly, the proposed phase identification models are compared for the read German DN
\item Secondly, we quantify the impact of measurement proximity on phase identification metrics. We observe that for a DN partitioned into zones, the estimation quality deteriorates as the measurement reference selected is farther away from the zone where the consumer is located.
\item Thirdly, the impact of measurement errors on PCI is assessed. For 1\% accuracy class measurements, an estimation accuracy exceeding 98.6\% is achieved for a DN with 646 nodes.
\item Most European DNs are four-wire system. In our last case study, we quantify the impact of the neutral conductor model on phase estimation accuracy. We observe that if the neutral conductor is not modeled, then a pessimistic estimation is achieved. Based on the knowledge of the authors, the neutral conductor impact assessment is done for the first time.
\end{itemize}
\end{itemize}

The paper is organized as follows. 
Section~\ref{section2} presents the German DN case of low observability and the need for enhanced phase information for future DN operation.
Section~\ref{section3} outlines the different modeling steps used for generating synthetic data used for phase connectivity identification.
Section~\ref{section4} presents the methodology, and 
Section~\ref{section5} presents the consensus algorithms used for phase identification.
Section \ref{section6} presents the four numerical case studies, and
section~\ref{section7} concludes the paper.

\pagebreak

\section{Low observability in DN: the German case}
\label{section2}

The low voltage grid serves households and small consumers connecting at 230 V or 400 V  \cite{link1}. German authorities have decided to implement {an optional} smart meter (SM) roll-out {as the information security standards need to be adjusted}. Presently, less than 5\% of residential customers are equipped with SMs \cite{link2, link526}. Thus, the smart meter infrastructure is not widespread at a low-voltage level. The German Energy Industry Act requires that customers with a yearly consumption of over 6000 kWh are provided with smart measurement systems (when technically possible) \cite{link998,link999}. The requirement also applies to generators with an installed capacity above 7 kW. However, these requirements leave the majority of German households unaffected. The lack of sufficiently granular metering equipment at the household level is currently a barrier for implementing imbalance sensitive flexibility activation for solving DN issues.

\subsection{Metering of German DN}

Mitnetz Strom is one of the largest regional distribution system operators in Eastern Germany and is responsible for supplying electricity to 2.2 million {electricity consumers}. The grid area of Mitnetz Strom covers an area of 30,804 km$^2$ and is characterized by rural conditions with a high share of renewables.
{
The installed capacity of renewable energy reached an all-time high of more than 10,000 MW (more than 64,0000 plants) in
2021. This development was spurred primarily by rapid growth in solar energy, as the number of photovoltaic installations
increased by more than 17 percent. 
\cite{link222}.}

{In Germany, the metering and DSO roles are decoupled. The Meter Point Operator (MPO) is responsible for the installation, operation, data gathering, and maintenance of energy meters. Note, in many locations, system operators also perform as MPO. However, the electricity consumer could opt for an independent MPO. This is in accordance with the § 43 German MsbG (measuring point operation law).}
In principle, also a third party can be commissioned as a meter operator with the operation of the metering point on the free market \cite{link3A1}.
%
Furthermore, there is presently no general legal obligation to share grid-relevant information obtained from smart meters with the DSO.
{Due to these constraints, DSO needs to request for receiving historical data thus cannot be utilized for short-term grid operation, congestion mitigation, etc (due to delays in DSO making a request and receiving the measurement data). Thus, observability in DNs are limited not only by SM penetration level but also by data-sharing policies targeting system operators.}

\subsection{Roadmap of meter rollout}
\begin{figure*}[!htbp]
	\center
	\includegraphics[width=0.85\linewidth]{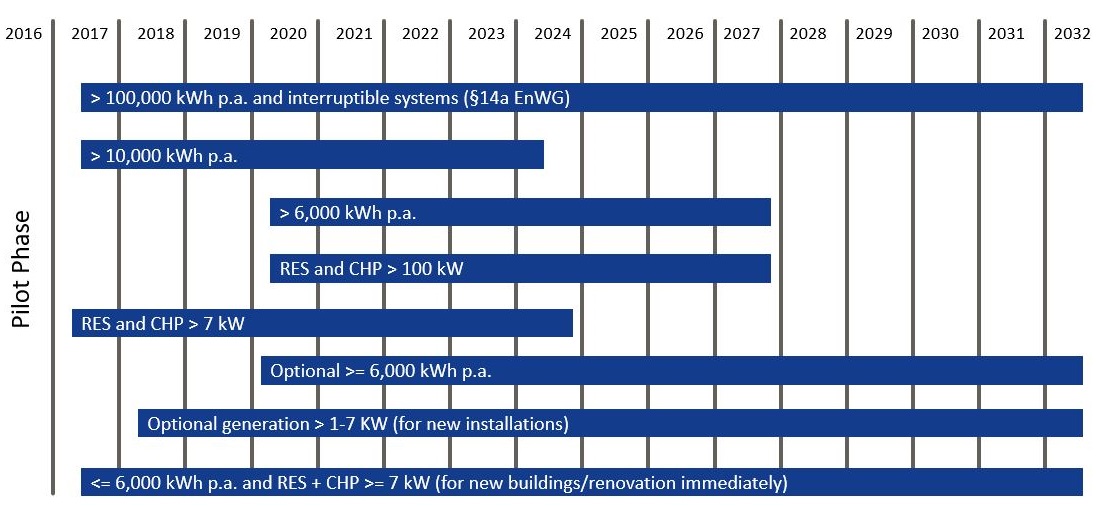}
	\vspace{-2pt}
	\caption{\small{Rollout Plan for smart meters in Germany by 2032 \cite{link4A2}} }
	\label{fig:pic1}
\end{figure*}

Fig. \ref{fig:pic1} shows the meter rollout phases in German. It was expected that by 2032, all German consumers are to be equipped with modern metering devices. (§ 29 para. 3 p.1 MsbG) compared to other countries, thus it will still take several years before the DSO has sufficient data from SMs at its disposal. 
{Complicating this schedule are also the}
legal issues concerning safety and privacy of smart meter operation and usage\footnote{On 20 May 2022, the Federal Office for Information Security (BSI) withdrew the general ruling of 7 February 2020 on the determination of technical feasibility pursuant § 30 MsbG (so-called market declaration on the rollout of smart metering systems) with effect for the past. In addition, the BSI issued a general ruling pursuant to §19 (6) MsbG in which it determined that the use and installation of smart metering systems available on the market do not pose any significant risks.}. The continued operation and installation of smart metering systems as defined by the Act by the MPOs is thus still possible. However, there is no longer an obligation to install them \cite{link5A3}. This makes the need for DN parameter estimation and mechanisms to enhance observability even more crucial for ensuring the operational integrity of the network.

Mitnetz Strom has invested a total of 19 million euros in 2022 in the conversion to digital local network stations.  The plan is to install a total of 226 digiONS in the year 2022. By 2026, up to 30 percent of the transformer stations and cable distributors in the network area are to be digitally equipped or retrofitted with the corresponding metering technology. Measured secondary substations are an important component in the digital transition. They ensure better controllability and transparency of medium and low-voltage grids, which directly benefits the implementation of the energy transition and security of supply. Increasing feed-in from renewable energies, rising demand for charging power for electro-mobility, extreme weather conditions that endanger the energy supply, especially in areas with overhead lines - the reasons for the digital monitoring and control of electricity grids are many and have one goal: \textit{security of supply}.


\subsection{Demo network for EUniversal}

In EUniversal, Mitnetz Strom is testing the use of flexibility services and markets and is leading the German demonstration together with the parent company E.ON SE \cite{link6}. The German demonstration tries to combine principles of the German mandatory process Redispatch 2.0 \cite{link888} and a market-based approach to mitigate grid constraints in a cascaded operation across multiple voltage levels. The goal is to provide DSOs with access to flexibility from grid customers across the LV/MV level for their active system management. To this end, the EUniversal consortium is testing various optimization algorithms with the aim of minimizing activation costs while ensuring the secure operation of the grid and is developing concepts for grid state estimation of smart grids, of which the first interim results and experiences will be presented.
In the German demo of EUniversal, Mitnetz Strom and its partners are investigating the use of flexibility markets in low voltage grids for congestion management and voltage maintenance. An attempt is being made to develop an iterative procedure that will prevent new congestion from occurring when flexibility is activated.

\subsubsection{Network features and meter placement}
The network considered for numerical evaluation is a typical low-voltage network in Mitnetz's network area in a small town in Eastern Germany. There are already some flexible plants, but the penetration with them is not yet significant.
Due to application cases, certain locations in the LV grid are particularly interesting when it comes to equipping with measurement technology. In particular, the end of the feeder is often an important indicator for the evaluation of potential voltage band violations, while the current at the beginning of the feeder is important for the thermal constraints in the network. Unfortunately, these points are often not available in practice due to ownership issues and the partially pronounced building development in the localities. Therefore, cable distribution cabinets were selected and equipped with measurements. 
For EUniversal devices are a bundle of voltage and current sensors (Rogowksi coils), gateway, and power supply. 











\subsection{Need for phase information}
LV DN topology identification is essential for efficient network operation, monitoring, and control. This also assists in planning the phases for new resources connected to the DN. Next, there is a real issue Mitnetz Strom faced due to the connection of 8 out of 9 electric vehicle chargers was made using a single-phase ($1-\phi$). This led to thermal limit violations in that particular phase. With the topology identification, the phase connections were optimized.
According to DIN ISO 50160, \cite{link7}, limits of voltage deviations are defined up to $\pm 10\%$. LV networks are asymmetrically biased by 1-$\phi$ loads. Unbalanced new loads, such as electric fans{, heat pumps,} and unbalanced charging {could} amplify this effect. Unbalanced loaded DNs lose a substantial amount of their power transmission capability. In a research project for an automatic phase switch in EV charging showed the effects in charging for $1-\phi$ or two-phase connected EVs or hybrids in the grid. 
{These findings presented here are also applicable}
to $1-\phi$ heaters, inverter interfaced PV, storage, and other unbalanced loads. 

\begin{figure}[!htbp]
	\center
	\includegraphics[width=0.6\linewidth]{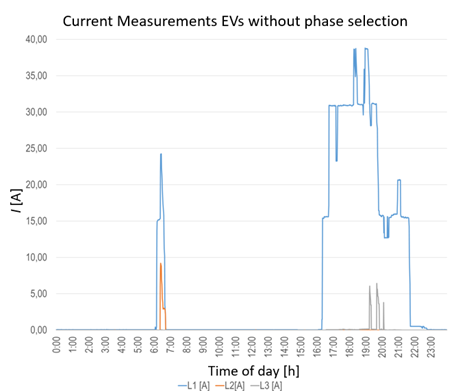}
	\vspace{-1pt}
	\caption{\small{Line loading for unplanned EV charger placement}}
	\label{fig:pic2}
\end{figure}
\begin{figure}[!htbp]
	\center
	\includegraphics[width=0.6\linewidth]{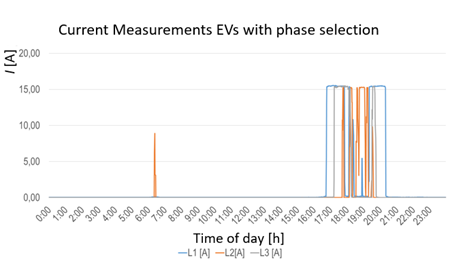}
	\vspace{-1pt}
	\caption{\small{Line loading after phase redistribution }}
	\label{fig:pic3}
\end{figure}

In the further evaluation of this field test, it is shown that with the average existing charging capacity, the penetration rate with EVs until a line limit is violated increased from 16\% to 47\% with symmetrical utilization {of DN with approximately balanced phases}. 
In the case of flexibility markets, this means the possibility of using flexibility decreases for ensuring grid integrity. 
{This underlines the importance of having accurate knowledge}
of the phase connections, as flexibility activation should not further increase imbalance. In theory, flexibility activation could also limit DN imbalance.
Mitnetz Strom and most DSO’s follow a passive way of identifying the phase connections. The accurate topology identification is performed manually in case of a follow-up on customer complaints on power quality. 
This motivates us to propose a scalable and robust phase identification mechanism using historical measurement data.

\pagebreak

\section{Synthetic data generation}
\label{section3}

In this section, we detail the steps we took for generating synthetic data for the DSO grid layout provided in DigSilent format. A digital twin is used for the process of phase and load placement.
Further, the neutral conductor modeling, noise injection, and zonal clustering model used in this work are elaborated {in this section}.
\subsection{DGS parser for network JSON}
A parser has been created to convert the DGS (\textbf{D}I\textbf{gS}ilent) \cite{manual2011digsilent} file format into a JSON file that is readable to the PowerModels script. The parser is derived from the GridCal python package \cite{gridcal}. The main difference between the two formats is that the DGS data format contains different classes with different information in a hierarchical structure, whereas the JSON file just requires information on the buses, branches, and devices in the grid.

\subsubsection{Bus information}
The DGS format relies on cubicle information, which can be seen as a connection point for the different elements in the grid. The cubicle information are the unique IDs given to each connection point, which are converted to simple grid ids starting from 1 and ending in the number of buses in the grid. The original ID is saved to allow for cross-checking data and for linking devices to the correct bus.

\subsubsection{Branch information}
The DGS branch format relies on cable type information, but the JSON file requires the values directly in per-unit (pu). Therefore, the \textit{r} and \textit{x} parameters (amongst other) needs to be converted from their ohmic values to pu of the total cable length. This requires a multiplication of the base impedance and the total length of the cable.


\subsubsection{Device information}
The DGS file format can contain detailed information on different loads and static/synchronous generators. The JSON file format only has information on the devices. Therefore, the parser extracts the relevant P and Q data from each device in the DGS file and compiles it as a separate entity in the device file. Additional information included is bus ID, PV size, and connected phase.

\subsubsection{Switches}
A crucial element of the parser is removing the switches that connect branches to substations and cabins. The switches are removed as they add numerical complexities when calculating the admittance matrix in PowerModels \cite{8442948}. Setting the \textit{r} and \textit{x} values to zero can lead to infinity values during the admittance calculation, but setting it to a very low value leads to inefficiencies when running the power flow. Removal of the switches not only removes numerical complexities but, since switches make up ~7\% of the branches in the system, by removing the extra nodes the computation time of the simulation significantly increases {(in the order of a 10\%)}. 

Once the switches have been removed, the IDs should be renumbered. This also serves the purpose of removing empty nodes in the grid, which has a positive effect on the computation time of the simulation as the admittance matrix only contains non-zero components, which helps reduce its size. This also tidies up the network data, making it easier to view from a simulation perspective.


\subsection{Mitnetz Strom DN and metadata}
\begin{figure}[!htbp]
	\center
	\includegraphics[width=5.5in]{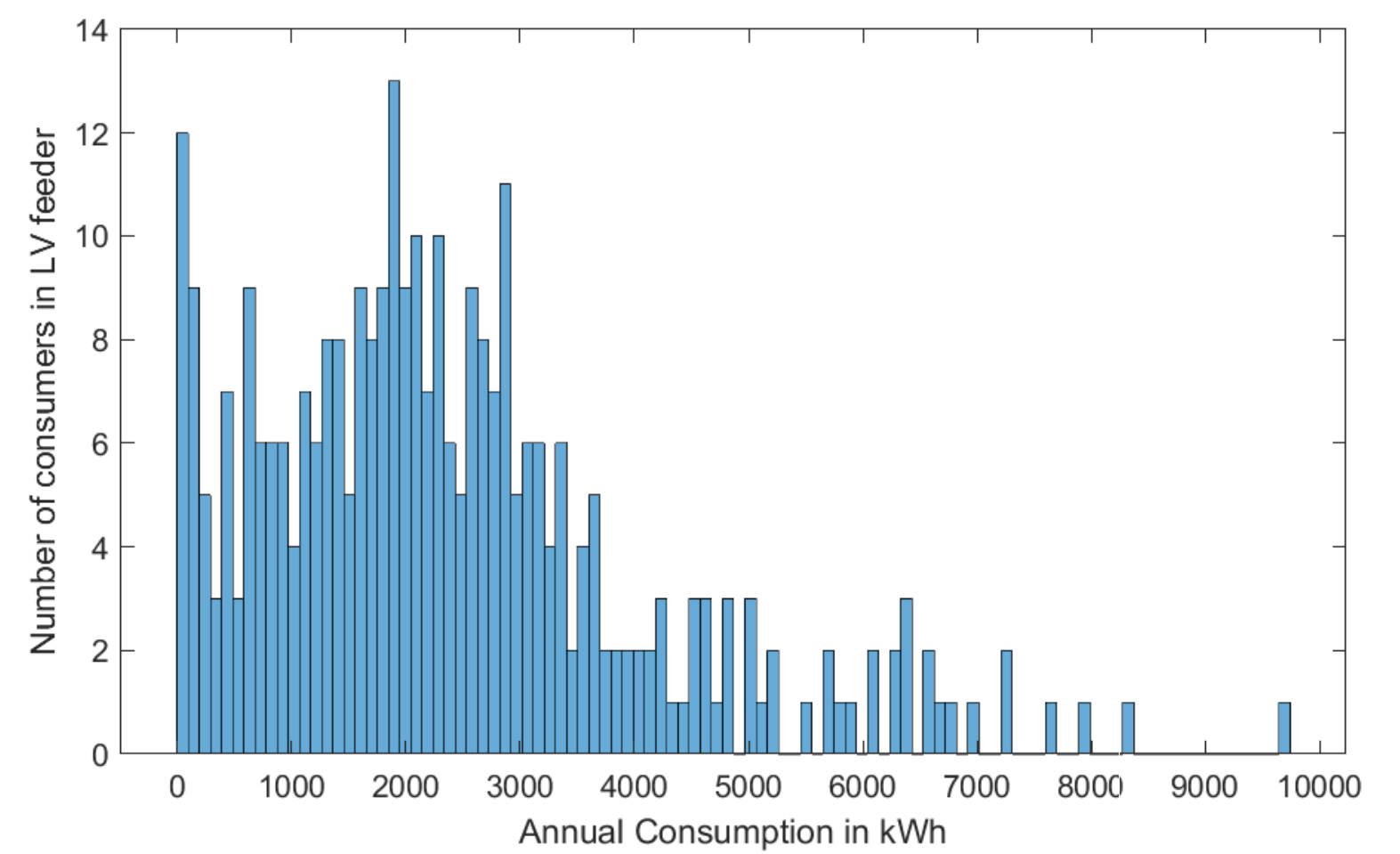}
	\vspace{-2pt}
	\caption{\small{Metadata for annual kWh consumption of 331 consumers in the test LV DN considered. {Note that 94.5\% of consumers have an annual consumption of lower than 6000 kWh for the test DN.}}}
	\label{fig:distkwh}
\end{figure}

Along with the grid data, metadata on the loads in the grids are also provided in a separate file. This metadata contains details associated with different consumer devices in the grid. The information includes a node number to connect it to the grid data, load type (e.g. household, PV, CHP, etc.), the annual energy consumption, single or three-phase connection, and the available active and reactive power output (if applicable). The information is compiled and appended to the device's JSON file. Fig. \ref{fig:distkwh} shows the spread of the annual cumulative energy consumption of loads connected to the {test DN used in this work}.

Note that the exact phase connection of single-phase loads is not known. The goal of this paper is to present a framework for identifying the phase connection of such single-phase consumers.

\subsection{{Randomized} phase mapping}

DSOs actively try to balance the phases so that the load distribution is fairly balanced.
Madeira island case study in \cite{hashmi2020towards} and DSO questionnaire in \cite{hashmi:tel-02462786} details the phase assignment procedure of a DSO.

In order to generate synthetic data, randomized load mapping is used. For a single phase load with different levels of annual kWh consumption, a phase is randomly selected from phases A, B, and C with equal likelihood.

\begin{figure}[!htbp]
	\center
	\includegraphics[width=5.1in]{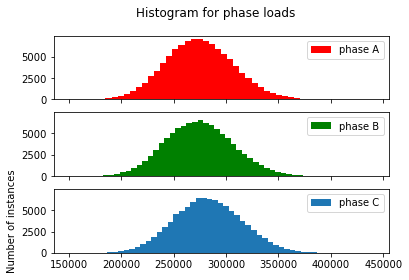}
	\vspace{-2pt}
	\caption{\small{Phase load distribution of three-phase distribution network for 100000 phase mapping scenarios.}}
	\label{fig:phaseload}
\end{figure}

The randomized phase mapping is evaluated based on the sum of the absolute error in phase load (AEPL), which is given as
\begin{gather*}
    AEPL = \frac{1}{3}\frac{1}{D}\sum_{i=1}^D\sum_{\phi\in\{A,B,C\}} |L_{\phi}^D - \bar{L}^D|,
\end{gather*}

where $D$ denotes the number of Monte Carlo phase mapping scenarios, and $\bar{L}^D$ denotes the mean load in all the phases and is given as $\bar{L}^D =  \sum_{\phi\in\{A,B,C\}} L_{\phi}^D/3$.

\begin{figure}[!htbp]
	\center
	\includegraphics[width=5in]{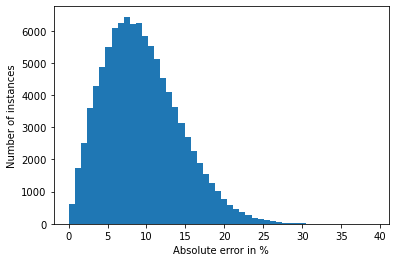}
	\vspace{-2pt}
	\caption{\small{Sum of the absolute value of the difference of phase load and mean load of all the phases.}}
	\label{fig:error}
\end{figure}

Using 100000 Monte Carlo simulations for phase mapping, we observe that randomized phase mapping performs fairly well, with a maximum and mean per phase load deviation of 30\% and 9.5\% with respect to the mean load met by the three phases in the worst case (Fig. \ref{fig:error}).
Fig. \ref{fig:error} shows the distribution of the phase load errors while performing randomized phase mapping. 



Thus, in this work, we utilize randomized phase mapping for data generation for evaluating our proposed probabilistic phase identification mechanism.

\subsection{Neutral modelling {for four-wire DN}}
\begin{figure}[!htb]
    \centering
    \includegraphics[width=12cm]{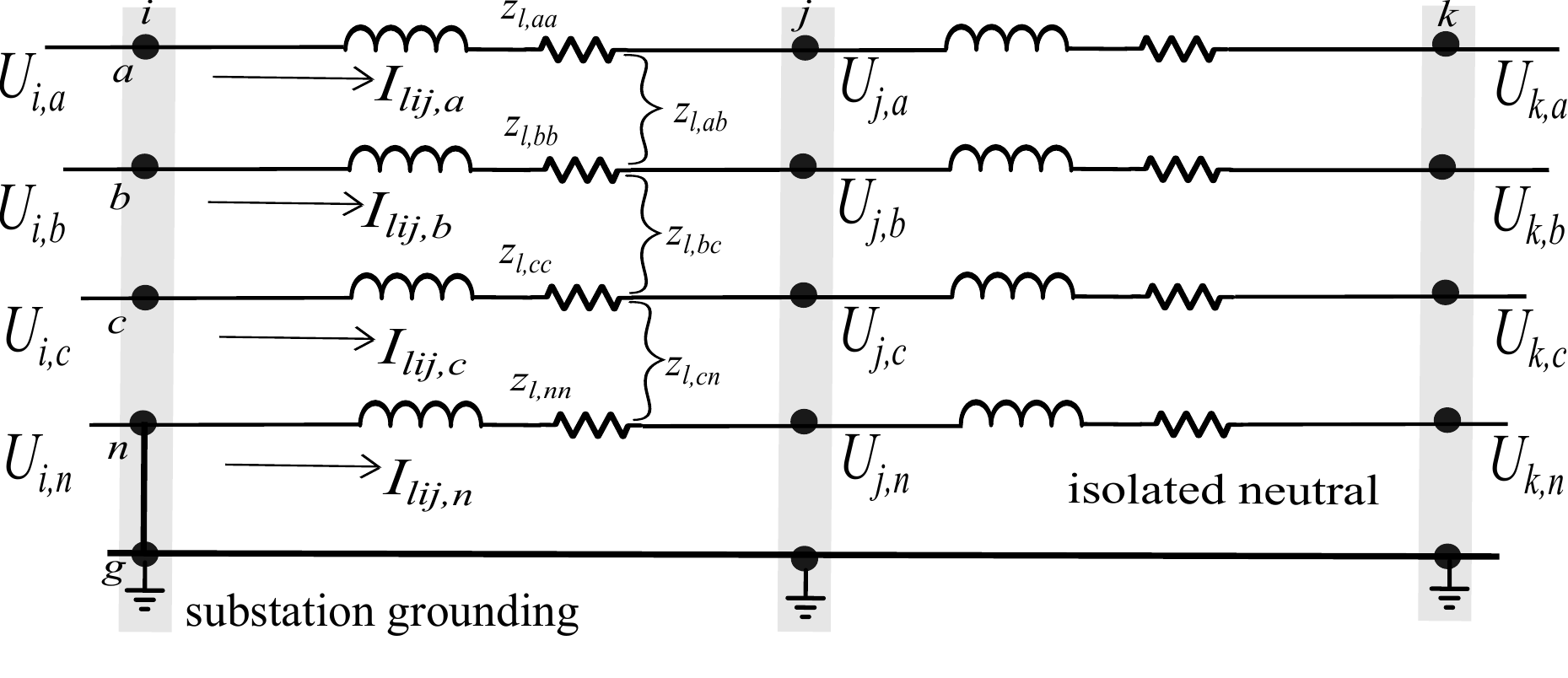}
    \caption{Isolated neutral model of distribution network}
    \label{fig_EU}
\end{figure}

European LV DN is usually different from the North-American one with a larger size distribution transformer with multiple low voltage feeders supplying a large number of consumers per transformer. The German low voltage feeders normally follow a four-wire three-phase configuration with single-grounded neutral~\cite{en14051265}. {Such systems are usually reduced to three-wire equivalent using Kron's reduction\cite{Kersting2001b}. In Kron's reduction, it is assumed that the neutral is grounded multiple times and for a perfectly grounded neutral\footnote{A perfectly grounded neutral refers to grounding resistance of zero ohms. Typically the grounding resistance is $\approx$ 5 ohms which leads to a small voltage drop. In this work, we assume perfectly grounded neutral.}, the neutral voltage equals zero.} However, for the DN considered in this paper, this assumption is not true as {the neutral is isolated from consumer grounding and is grounded only at the sub-station~(See Fig.~\ref{fig_EU}). The inclusion of sparsely grounded neutral in modeling can be done by taking an exact four-wire model with four-wire power flow solvers or reducing it to a three-wire equivalent and solving by using the three-wire solvers.}  In~\cite{geth2022computational} a new reduction method is proposed for sparsely grounded European LV feeders so that the impact of neutral is represented as equivalent as a four-wire model without the necessity of carrying around extra variables and measurements. In such reduction, the 4$\times$4 impedance matrix is transformed to 3$\times$3 matrix is given in \eqref{eq:impedancematrix}.



\begin{equation}
\text{Impedance matrix}~ =
\begin{bmatrix} 
\zijksAA - \zijksNA - \zijksAN + \zijksNN  ~~&\zijksAB  - \zijksNB - \zijksAN + \zijksNN  ~~& \zijksAC  - \zijksNC - \zijksAN + \zijksNN  \\
\zijksBA - \zijksNA - \zijksBN + \zijksNN  ~~&\zijksBB - \zijksNB - \zijksBN  + \zijksNN   ~~& \zijksBC - \zijksNC - \zijksBN + \zijksNN \\
\zijksCA - \zijksNA - \zijksCN + \zijksNN  ~~& \zijksCB - \zijksNB - \zijksCN+  \zijksNN  ~~& \zijksCC - \zijksNC - \zijksCN  + \zijksNN  \\
 \end{bmatrix} 
 \label{eq:impedancematrix}
\end{equation}

\vspace{5mm}
  
In \eqref{eq:impedancematrix}, $\zijksAA$ is self-impedance of the phase $a$ of the branch $l$, $\zijksNN$ is self-impedance of the neutral of the branch $l$, and so on. Similarly, $\zijksAB$ is the mutual-impedance between phase $\symPhaseA$ and $\symPhaseB$ of branch $l$. This transformation is exact and eliminates the error introduced by Kron's reduction in three-phase DN modeling for sparsely grounded system~\cite{geth2022computational, Koirala2019}. Furthermore, a minor boost in computation time is achieved {compared to the exact four-wire model} as the necessity of carrying extra variables for neutral voltage is removed. This reduction is more relevant as the measured voltages in {German demo-grid} are also phase-to-neutral. Interested readers are guided to~\cite{geth2022computational} for details about the transformation.


\subsection{Metering noise injection model}
Prior works \cite{pi2:pappu2017identifying, pi32:heidari2021phase, pi18:xiaoqing2018phase, pi9:hoogsteyn2022low} consider smart meter measurement error based on the accuracy class of metering infrastructure.
Frequently, the measurement error is considered using Gaussian noise. Further, the measurement accuracy is considered to hold true for three sigma of the times, which corresponds to 99.7\% of total instances.
The standard deviation of the Gaussian noise is related to the tolerance $\tau$ of the measuring device. 
The noisy measurement is given as
\begin{equation}
    \hat{Z} = Z \times \texttt{Norm}(1, \tau/3),
    \label{eq:measurmenterror}
\end{equation}
where $Z$ denotes the true measurement, $\texttt{Norm}(\mu,\sigma)$ denotes a sample of a normal distribution with mean $\mu$ and standard deviation $\sigma$.
In order to evaluate the impact of measurement noise, 1000 Monte Carlo simulations are considered.

\subsection{Zonal clustering of Mitnetz Strom DN}
Identifying the zones of an LV DN can be helpful to the DSO in planning the flexibility needs of a network. Due to the large numbers of DN feeders, a standardized approach to divide zones based on electrical and/or geographical distances is deemed essential \cite{https://doi.org/10.48550/arxiv.2207.10234}.
In this section, the summary of the clustering framework to identify the best-suited LV DN zonal partition using electrical distance as a measure is presented, which is explained in detail in \cite{https://doi.org/10.48550/arxiv.2207.10234}.
\textcolor{black}{This zonal partition method uses an incidence matrix-based measure, which can be obtained with the help of spectral decomposition of the admittance matrix. The adequate number of zones is obtained based on the maximization of silhouette score \textcolor{black}{while considering the desired number of clusters}}.

The zonal partition divides nodes $\mathscr{N}$ into $c \in\{1,...,C\}$ clusters.
The spectral clustering proposed in~\cite{ding2018clusters, sanchez2014hierarchical} for the creation of zones or network reduction is used for zonal partition.
A double stochastic matrix is formed, which is a special type of Markov matrix where not only each row but also each column add to 1. 
For this transformed matrix, all eigenvalues are real and smaller than or equal to 1, with one eigenvalue exactly equal to 1~\cite{mourad2012spectral}.
For identifying $C$ partitions in a graph, the $C$ highest eigenvalues and corresponding orthonormal eigenvectors are identified.
The eigenvector matrix of the order $N \times C$ is used for DN partitioning, in effect reduces the dimensionality of the problem.
$k$-means clustering is used to partition the spectral data points.
The goodness of a cluster is measured using the mean silhouette index of the network cluster.
The silhouette coefficient of a node is a confidence indicator of its association in a group~\cite{ding2018clusters, scarlatache2012using}.

\subsection{Power flows for synthetic data}

Power flow equations translate the load information of consumers to the nodal voltage and nodal currents when the network topology and impedances are known using the first equations. Three-phase unbalanced power flow equations were used to create the pseudo-measurement point based on the given load data and network topology. Open-source power flow solver of \texttt{PowerModelsDistribution.jl} was used for creating such pseudo measurement points~\cite{Fobes2020}.

\pagebreak

\section{Phase identification methodology}
\label{section4}
Using the synthetic data generated in the previous section, we develop correlation based voltage matrices used for consensus algorithm base PCI algorithms in the next section.

\subsection{Notation}


A three-phase distribution network (DN) consists of phases denoted as $\phi \in \{A,B,C\}$. The DN consists of branches, nodes, loads, and generators.
A DN is represented as a directed graph by $<\mathscr{N},E>$, where $\mathscr{N}$ denotes the set of nodes in all the phases and $E$ denotes the set of branches connecting a pair of nodes. 
For each node $i$ in the phase $\phi$ at any time $t$, have two variables: (i) voltage magnitude denoted as $V_{{\phi},i,t}$ and phase angle denoted as $\theta_{{\phi},i,t}$. The voltage phasor at a node and phase is governed by power injections.
The branch denoted as $(i,j) \in E$ is characterized by line admittance denoted as $Y_{\phi,ij}$. The line admittance governs power flow and line losses.
$\mathscr{N}_d \subset \mathscr{N}$ denotes the nodes with loads connected. For these nodes, the active and reactive power is given as $P^d_{{\phi},{i,t}}$ and $Q^d_{{\phi},{i,t}}$.
$\mathscr{N}_g \subset \mathscr{N}$ denotes the nodes with generators connected, have active and reactive power generation denoted as $P^g_{{\phi},{i,t}}$ and $Q^g_{{\phi},{i,t}}$.
The time $t$ is sampled hourly, and its range is given as $t \in \{1,..,T\}$.

\begin{figure}[!htbp]
	\center
	\includegraphics[width=5.2in]{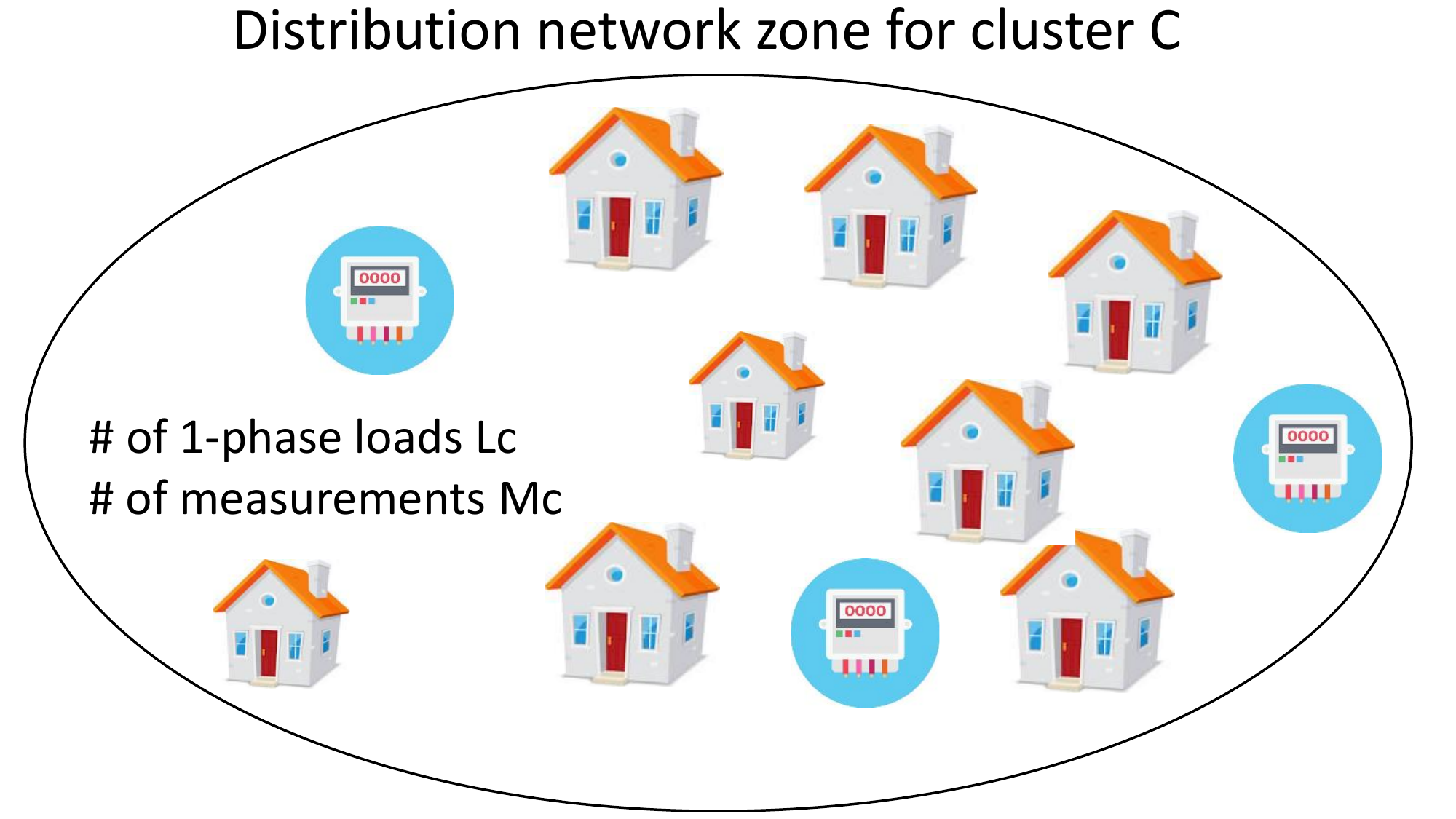}
	\vspace{-2pt}
	\caption{\small{A distribution network cluster with measurement points and single-phase loads with unknown phase connectivity.}}
	\label{fig:diag1}
\end{figure}

The DN is clustered into $c \in \{1,...,C\}$ clusters.
A cluster $c$ consists of $M_c$ number of three-phase reference measurement points, $L_c$ number of single-phase consumers and $N_c$ are the number of nodes {present} in that cluster.
{
We assume that all the reference measurements are aligned and there are no synchronization delays considered in this work.
A stylized representation of a zone with $1-\phi$ consumers and $3-\phi$ reference measurement points are shown in Fig. \ref{fig:diag1}.}
DN clustering is performed such that $N_i \subset \mathscr{N}$ and $N_i \cap N_j = \emptyset, ~ \forall i \neq j$.
The set of nodes where $M_c$ measurements are placed is denoted as $i_c^M$.
The set of nodes where $L_c$ single phase consumers are located is denoted as $i_c^L$.
For a vector parameter $K$, $\bar{K}$ is the mean value, and $|K|$ denotes its absolute value.
For a vector $K$, $\mathscr{C}(K)$ denotes its cardinality.
$\mathbbm{1}{(\text{condition})}$ returns 1 if the condition is true.

\subsection{Correlation based metrics}
In this work, we utilized voltage magnitude time series as a parameter for estimating phases of the single-phase consumers in each of the clusters.
Each of the measurement points is utilized for estimating the phases. Thus, a unique reference matrix is created using the phase voltage magnitudes at the measurement node. For cluster $c$, measurement $i_c^M(j)$ where $j\in \{1,..,M_c\}$, the reference voltage matrix is given as

\begin{equation}
\small
V^{\text{ref}}_{i_c^M(j)} =
{\begin{bmatrix}
V_{A,i_c^M(j), 1} & V_{B,i_c^M(j), 1} & V_{C,i_c^M(j), 1}\\
V_{A,i_c^M(j), 2} & V_{B,i_c^M(j), 2} & V_{C,i_c^M(j), 2}\\
:& : & :\\
:& : & :\\
V_{A,i_c^M(j), T} & V_{B,i_c^M(j), T} & V_{C,i_c^M(j), T}\\
\end{bmatrix}} 
\end{equation}

The dimension of $V^{\text{ref}}_{i_c^M(j)}$ is $T\times 3$.
Note time $t$ is hourly, therefore, $\mathscr{C}(t\in\{1,..,T\})=T$.

The single phase consumer nodal voltage time series in cluster $c$ forms a column of the matrix denoted as $V_c^L$, and given as

\begin{equation}
\small
V^{\text{L}}_{c} =
{\begin{bmatrix}
V_{\phi,i_c^L(1), 1} & V_{\phi,i_c^L(2), 1} &..& V_{\phi,i_c^L(N_c), 1}\\
V_{\phi,i_c^L(1), 2} & V_{\phi,i_c^L(2), 2} &..& V_{\phi,i_c^L(N_c), 2}\\
:& : &..& :\\
:& : &..& :\\
V_{\phi,i_c^L(1), T} & V_{\phi,i_c^L(2), T} &..& V_{\phi,i_c^L(N_c), T}\\
\end{bmatrix}} 
\end{equation}

The dimension of $V^{\text{L}}_{c}$ is $T\times N_c$.
The connection phase of $1-\phi$ consumers are assumed to be not known.

{For the calculation of correlation between reference $3-\phi$ voltage and $1-\phi$ consumer voltage time series,}
Pearson correlation\footnote{The Pearson correlation between two vectors $X$ and $Y$ is given as
\begin{equation}
    \rho({X,Y}) = \frac{\sum (X-\bar{X}) (Y-\bar{Y})}{  \sqrt{ \sum (X-\bar{X})^2 \sum (Y-\bar{Y})^2 } }
\end{equation}} is utilized. 

Three models are presented for generating metrics for phase identification using a consensus algorithm.
These three metrics are described next.
\subsubsection{J1: correlation of voltage time series}
The correlation matrix for measurement $i_c^M(j)$ is denoted as
\begin{equation}
\small
\begin{split}
& \rho_{J1}^{c,i_c^M(j)} = \\
&{\begin{bmatrix}
\rho(V_{i_c^L(1)}, V_{A,i_c^M(j)}),  & \rho(V_{i_c^L(1)}, V_{B,i_c^M(j)})&\rho(V_{i_c^L(1)}, V_{C,i_c^M(j)})\\
:& : & :\\
:& : & :\\
\rho(V_{i_c^L(N_c)}, V_{A,i_c^M(j)}),  &
\rho(V_{i_c^L(N_c)}, V_{B,i_c^M(j)})&\rho(V_{i_c^L(N_c)}, V_{C,i_c^M(j)})\\
\end{bmatrix}} 
\end{split}
\end{equation}
Note that the voltage time series is denoted by dropping $t$ in the notation. 
For instance, $V_{i_c^L(w)}$ denotes the voltage time series for $t \in \{1,..,T\}$ for single phase consumer located at node id $i_c^L(w): w \in \{1,..,N_c\}$.
For single-phase consumers with unknown phases, the phase notation is also dropped to avoid confusion.

\subsubsection{J2: Salient features with voltage difference time series}
Salient features in voltage time series could help in improving phase identification. 
The use of salient features has been explored in \cite{pi3:ni2017phase, pi6:vycital2019phase, pi5:xu2016phase}. In this work, we utilize the difference matrix and a zonal voltage fluctuation threshold for identifying the salient features.
The difference matrix for the reference voltage matrix for cluster $c$ and measurement $i_c^M(j)$ is given as
\begin{equation}
    \Delta V^{\text{ref}}_{i_c^M(j)} = \Big[V_{\phi,i_c^M(j), t+1} - V_{\phi,i_c^M(j), t}~~ \forall~ t, \forall~\phi \Big].
\end{equation}
The dimension of $\Delta V^{\text{ref}}_{i_c^M(j)}$ is $(T-1)\times 3$.

$\beta_c$ denotes the voltage change threshold for cluster $c$.

The salient features are extracted using the $\Delta V^{\text{ref}}_{i_c^M(j)}$ matrix as
\begin{equation}
    i^{\text{salient}}_{c,i_c^M(j)} = \arg \mathbbm{1}\Big({|\Delta V^{\text{ref}}_{i_c^M(j)}(t)| > \beta_c~ \forall t\in \{1,..,T-1\}}\Big).
\end{equation}
The voltage difference matrix for the connected load matrix is denoted as 
\begin{equation}
    \Delta V_c^L = \texttt{diff}(V_c^L),
\end{equation}
where \texttt{diff} operator finds the difference of adjacent rows. The dimension of $\Delta V_c^L$ matrix is $(T-1)\times N_c$.
The new reference and load matrix extracts the rows with salient features in the reference matrix and are given as
\begin{gather}
   \Delta V^{\text{ref, J2}}_{i_c^M(j)} = \Big[  \Delta V^{\text{ref}}_{i_c^M(j)}(i^{\text{salient}}_{c,i_c^M(j)}, :) \Big],\\
   \Delta V^{\text{L, J2}}_{c} = \Big[  \Delta V_c^L(i^{\text{salient}}_{c,i_c^M(j)}, :) \Big],
\end{gather}
The correlation matrix with salient features using the voltage difference as a metric is given as
\begin{equation}
\small
\begin{split}
&\rho_{J2}^{c,i_c^M(j)} = \\
&{\begin{bmatrix}
\rho(\Delta V^{\text{L, J2}}_{c}(1), \Delta V^{\text{ref, J2}}_{A,i_c^M(j)})  & ..&\rho(\Delta V^{\text{L, J2}}_{c}(1), \Delta V^{\text{ref, J2}}_{C,i_c^M(j)})\\
:& : & :\\
:& : & :\\
\rho(\Delta V^{\text{L, J2}}_{c}(N_c), \Delta V^{\text{ref, J2}}_{A,i_c^M(j)})  & ..&\rho(\Delta V^{\text{L, J2}}_{c}(N_c), \Delta V^{\text{ref, J2}}_{C,i_c^M(j)})\\
\end{bmatrix}} 
\end{split}
\end{equation}

\subsubsection{J3: Salient features with voltage magnitude time series}
Previously, we used the voltage difference as a metric for identifying the salient features. The salient features when projected onto the voltage magnitude would require the previous time stamp to capture the voltage change trajectory. This trajectory captured will improve the correlation-based metric we are utilizing for phase identification.
The new salient feature matrix is given as
\begin{equation}
    i^{\text{sal, plus}}_{c,i_c^M(j)} = \texttt{unique}(\big[i^{\text{salient}}_{c,i_c^M(j)}, i^{\text{salient}}_{c,i_c^M(j)} +1  \big]), 
\end{equation}
with \texttt{unique} operator finding unique time stamps, considering there could be repetitions that will be eliminated.

The new reference and load voltage matrix are given as
\begin{gather}
    V^{\text{ref, J3}}_{i_c^M(j)} = \Big[  V^{\text{ref}}_{i_c^M(j)}(i^{\text{sal, plus}}_{c,i_c^M(j)}, :) \Big],\\
   V^{\text{L, J3}}_{c} = \Big[  \Delta V_c^L(i^{\text{sal, plus}}_{c,i_c^M(j)}, :) \Big].
\end{gather}

The correlation matrix for model J3 is given as
\begin{equation}\begin{split}
&\rho_{J3}^{c,i_c^M(j)} = \\
&{\begin{bmatrix}
\rho(V^{\text{L, J3}}_{c}(1),  V^{\text{ref, J3}}_{A,i_c^M(j)}),  & ..&\rho(V_{i_c^L(1)}, V^{\text{ref, J3}}_{C,i_c^M(j)})\\
:& : & :\\
:& : & :\\
\rho(V^{\text{L, J3}}_{c}(N_c), V^{\text{ref, J3}}_{A,i_c^M(j)}),  &
..&\rho(V_{i_c^L(N_c)}, V^{\text{ref, J3}}_{C,i_c^M(j)})\\
\end{bmatrix}} 
\end{split}
\end{equation}

The dimension of $\rho_{J1}^{c,i_c^M(j)},\rho_{J2}^{c,i_c^M(j)}$ and $\rho_{J3}^{c,i_c^M(j)}  $ equals $N_c \times 3$.

\pagebreak

\section{Consensus algorithm}
\label{section5}

In Section \ref{section4}, we calculated correlation matrices using the voltage time series for measurement reference located at node $i_c^M(j), j \in \{1,..,M_c\}$.
Thus, with multiple measurement points in a cluster, independent phase identifications can be performed. These estimations can be taken into consideration using consensus algorithms to be presented in this section.


A consensus algorithm is a strategy that a group of agents use to agree with each other on what's true.
In a multi-sensor PCI scenario, there is just one true phase placement (ground truth), which is given as $P^{\text{true}}_c$. Each of the measurements used as a reference for models J1, J2, and J3 as metrics are used for estimating the true phases of single-phase consumers in cluster $c$.
The advantage of the consensus algorithm is that no one measurement point limits the PCI accuracy.
One of the most widely used consensus algorithms is in blockchain technology.
Consensus algorithms are widely used in state estimation \cite{con2:rana2017consensus, con3:soatti2016consensus, con4:xia2019distributed}.
In this work, we use consensus for phase identification in a distribution network.

\subsection{Na\"{i}ve phase identification}
The na\"{i}ve phase identification, {denoted as $S_0$}, uses only one of the $3-\phi$ reference measurement points, typically the substation measurement.
\begin{equation}
    P^{\text{est}, Jx, S_0}_{c} = \arg\max \rho_{Jx}^{c,i_\text{sel}},
\end{equation}
where $i_\text{sel}$ denotes the node id for the {reference measurement point.}
\textcolor{black}{Most literature on phase identification uses this na\"{i}ve model with substation time series measurement as the reference, see Tab. \ref{tab:phaselit}.}
\subsection{Majority rule}
The majority rule, {denoted as $S_1$}, is one of the most commonly used consensus algorithm. It identifies the most agreed-upon estimation. Earlier works such as \cite{maj1:goloboff2001methods, maj2:chappell2004majority} detail the applications of the majority rule in building consensus among agents (sensors).

We note that for a cluster $c$, measurement point located at node $i_c^M(j)$, and metric $Jx$, we can calculate $\rho_{Jx}^{c,i_c^M(j)}$. This correlation matrix is used for calculating the estimated phases, as
\begin{equation}
\begin{split}
    P^{\text{est}, Jx}_{c,i_c^M(j)} = \arg \max \rho_{Jx}^{c,i_c^M(j)},\\
    P^{\text{est}, Jx, S_1}_{c} = f_{S_1}\Big(P^{\text{est}, Jx}_{c,i_c^M(j)} ~ \forall j \in \{1,..,M_c\}\Big).
\end{split}
\end{equation}
The function $f_{S_1}$ calculates the majority among the estimated phases. Consider there are 7 measurement points in a cluster. For a node, consider 3 of the estimations that predict phase B, and 2 for phases A and C respectively. In this case, the majority rule predicts the phase to be estimated as phase B.

\subsection{Weighted measure}
Previously, for a majority rule-based consensus algorithm, we assumed all agents to be of equal importance (or weights). However, if we use the physical laws governing the system, we can calculate the weights for different agents. In phase identification, earlier works point out that measurement points in geographical proximity will have a greater voltage correlation among similar phases \cite{pi17:olivier2017automatic, pi30:short2012advanced}, see Tab. \ref{tab:phaselit}.
In this work, we use the correlation value as a weighing factor for calculating the estimated phase. 
A correlation value of 1 implies 100\% correlation.

\subsubsection{Correlation as a measure}

\begin{equation}
    \bar{\rho}^c_{Jx} = \sum_{k=1}^{M_c} \sum_{\phi=1}^3 \rho^{c,k}_{Jx}, 
\end{equation}
where $\bar{\rho}^c_{Jx}$ is $N_c\times1$ vector.
The normalized correlation coefficients are given as
\begin{equation}
    G^{Jx,S_2}_c = \frac{\sum_{k=1}^{M_c} \rho^{c,k}_{Jx}}{\bar{\rho}^c_{Jx}},
    \label{eq:gs2}
\end{equation}
where $G^{Jx,S_2}_c$ is $N_c\times3$ matrix.
The estimated phases are given as
\begin{equation}
    P^{\text{est}, Jx, S_2}_{c} =  \arg \max G^{Jx,S_2}_c.
\end{equation}

\subsubsection{Absolute value of correlation as a measure}

\begin{equation}
    \bar{\rho}^c_{Jx, \texttt{abs}} = \sum_{k=1}^{M_c} \sum_{\phi=1}^3 |\rho^{c,k}_{Jx}|, 
\end{equation}
where $\bar{\rho}^c_{Jx, \texttt{abs}}$ is $N_c\times1$ vector.
The normalized correlation coefficients are given as
\begin{equation}
    G^{Jx,S_3}_c = \frac{\sum_{k=1}^{M_c} |\rho^{c,k}_{Jx}|}{\bar{\rho}^c_{Jx, \texttt{abs}}},
    \label{eq:gs3}
\end{equation}
where $G^{Jx,S_3}_c$ is $N_c\times3$ matrix.
The estimated phases are given as
\begin{equation}
    P^{\text{est}, Jx, S_3}_{c} =  \arg \max G^{Jx,S_3}_c.
\end{equation}

\subsubsection{Maximum value of correlation as a measure}

\begin{equation}
    \bar{\rho}^c_{Jx, \texttt{max}} = \max_{k=1,..,M_c} \max_{\phi=1,2,3} |\rho^{c,k}_{Jx}|, 
\end{equation}
where $\bar{\rho}^c_{Jx, \texttt{max}}$ is $N_c\times1$ vector.
The normalized correlation coefficients are given as
\begin{equation}
    G^{Jx,S_4}_c = \frac{\max_{k=1,..,M_c} |\rho^{c,k}_{Jx}|}{\bar{\rho}^c_{Jx, \texttt{max}}},
    \label{eq:gs4}
\end{equation}
where $G^{Jx,S_4}_c$ is $N_c\times3$ matrix.
The estimated phases are given as
\begin{equation}
    P^{\text{est}, Jx, S_4}_{c} =  \arg \max G^{Jx,S_4}_c.
\end{equation}

Note that \eqref{eq:gs2}, \eqref{eq:gs3} and \eqref{eq:gs4} denotes element wise division of vector of length $N_c$.

\subsection{Metrics for phase identification models}
\subsubsection{Modelling accuracy}
Consider, the true phase information in a cluster is given as $P^{\text{true}}_c$.
The estimation accuracy is denoted as
\begin{gather*}
    \text{Estimation accuracy} = \frac{\text{number of correct phase estimation}}{\text{total number of single phase consumers}}.
\end{gather*}
The phase estimation accuracy for metric $Jx \in \{J1, J2, J3\}$, cluster $c$ and measurement $i_c^M(j)$ is given as
\begin{equation}
    A^{Jx}_{c, i_c^M(j)} = 100 \times \Big\{ 1- \frac{ \sum \mathbbm{1}\Big(  P^{\text{est}, Jx}_{c, i_c^M(j)}- P^{\text{true}}_c \neq 0 \Big) }{N_c} \Big\},
\end{equation}
where $P^{\text{est}, Jx}_{c, i_c^M(j)}$ denote the phase estimation vector for cluster input metric $Jx$, cluster $c$ and measurement $i_c^M(j)$.

The estimation accuracy for consensus algorithm $Sy$ is given as
\begin{equation}
    A^{Jx, Sy}_{c} = 100 \times \Big\{ 1- \frac{ \sum_{n=1}^{N_c} \mathbbm{1}\Big(  P^{\text{est}, Jx, Sy}_{c}- P^{\text{true}}_c \neq 0 \Big) }{N_c} \Big\},
\end{equation}

Since there are three base metrics denoted as $Jx \in \{J1, J2, J3\}$ and five consensus models (including the na\"{i}ve model) denoted as $Sy \in \{ S_0, S_1, S_2, S_3, S_4\}$, therefore, we evaluate in total 13 phase identification models {(the na\"{i}ve model, $S_0$, is performed for J1 only)}. In numerical results, we will compare the benefits and shortcomings of these models. Estimation accuracy averaged over Q Monte Carlo simulations are denoted as $\bar{A}^{Jx, Sy}_{c}$.

\subsubsection{Confidence factor for phase identification}
Note that models $S_1,S_3$, and $S_4$ provide coefficients that add up to 1 (node-wise). Thus, $G^{Jx,S_3}_c$ and $G^{Jx,S_4}_c$ can be used to indicate probabilities of phase estimation.
For $S_1$, the estimation probabilities can be calculated by dividing $P^{\text{est}, Jx, S_1}_{c} $ with the number of measurements in a cluster, $M_c$.

We define the confidence factor as the minimum distance between the factor associated with the correct phase and the maximum of the two incorrect phases, over all nodes in the cluster $c$.
For the model, $S_2$ we normalized the confidence factor with the range of variation of $G^{Jx,S_2}_c$.
The proposed confidence factor will provide us with additional information about the robustness of our phase estimation output.
Note, the proposed confidence factor can lie in the range $\in[-1,1]$ for $S_1, S_3$, and $S_4$. A confidence factor close to 1 implies very high confidence in our phase estimation output.
The confidence factor for measurement $k\in \{1,..,M_c\}$ and cluster $c$ is given as $F^{ Jx, S_y}_{c,k}$.
The confidence factor for a cluster $c$ for all Monte Carlo scenarios is given as
\begin{equation}
    \bar{F}^{ Jx, S_y}_{c} = \frac{1}{Q\times M_c} \sum_{q=1}^Q \sum_{k=1}^{M_c} F^{ Jx, S_y}_{c,k},
\end{equation}
where $w\in\{1,..,W\}$ denotes the Monte Carlo scenarios.

\subsubsection{Standard deviation with measurement error}
$Q$ Monte Carlo simulations are considered for minimizing the measurement error biases on phase estimation. For each of Monte Carlo iteration $q \in \{1,...,Q\}$, calculate the standard deviation of the measure used for calculating 
$P^{\text{est}, Jx, Sy}_{c}$, denoted as $D^{ Jx, S_y}_{c}$.

\pagebreak

\section{Numerical results}
\label{section6}
The numerical case study considers a German DN with 646 nodes and 331 loads connected to it. Out of 331 loads, 313 loads (94.6\% of total consumers) are single-phase loads. The phase connections are {widely unknown} to Mitnetz Strom. The objective of the case study is to assess the phase identification algorithm proposed in this work, {benchmarked over na\"{i}ve phase identification model}.
The selected DN is part of the demo network selected for evaluation in the EUniversal project.
Mitnetz Strom placed 53 $3-\phi$ measurement {devices} in the DN. These measurement points will be considered as references used for PCI.
The flexibility participants will be provided with a Home Energy Management System (HEMS) which will provide measurements of load and voltages at the point of common coupling. In the case studies, we assume the time series of voltage measurements of all single-phase users are known. In the real world, only measurements of consumers with HEMS will be available. 

There are four case studies performed in this paper. 
The first case study compares the 12 phase identification models on different phase mappings.
The second case study quantifies the impact of reference location in a DN on the phase identification metrics proposed in the work.
The third case study assesses the impact of measurement error at the reference and/or at the consumer location on phase identification metrics.
The last case study compares the phase identification metrics for DN with and without the neutral conductor model. As most European DNs are four-wire, it is crucial to quantify the impact.

Prior to case studies, we detail the {test DN} clustering results and the performance of the na\"{i}ve phase identification algorithm. The benefits of the proposed phase identification algorithms are compared to the na\"{i}ve model.

\subsection{Clustering of distribution network}
\begin{figure}[!htbp]
	\center
	\includegraphics[width=0.9\linewidth]{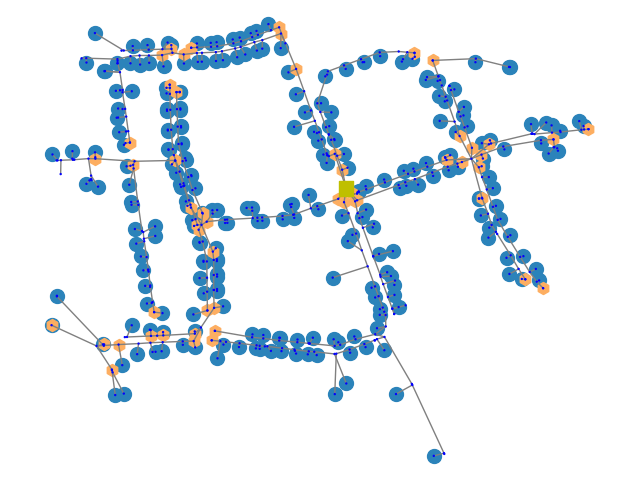}
	\vspace{-10pt}
	\caption{\small{Consumers (blue) and measurement points (orange) in the DN}}
	\label{fig:measurmentpts}
\end{figure}

Fig. \ref{fig:measurmentpts} shows the location of single-phase consumers and measurement points in the DN.
We apply the clustering algorithm for identifying the clusters in the DN.

\begin{figure}[!htbp]
	\center
	\includegraphics[width=0.9\linewidth]{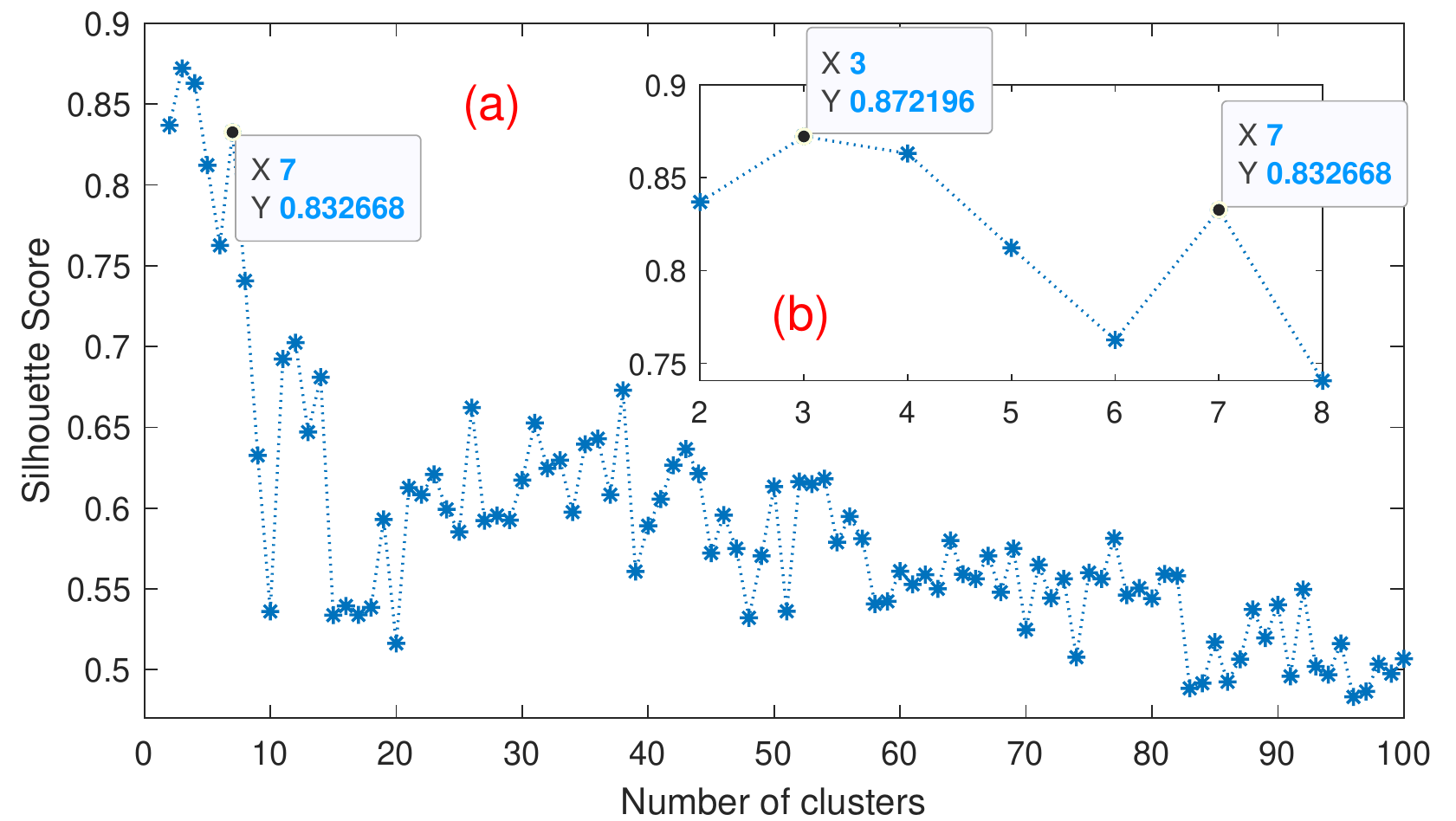}
	\vspace{-1pt}
	\caption{\small{Choosing the number of clusters of DN based on the silhouette coefficient. In (a) the variation of silhouette coefficient is plotted with increasing number of cluster. In (b) we zoom into the plot (a). Note the silhouette coefficient is maximum for 3 clusters, however, the best-suited number of clusters selected is 7 \cite{https://doi.org/10.48550/arxiv.2207.10234}.}}
	\label{fig:silhouette}
\end{figure}

Since the number of clusters to be formed is not clear, we utilize the silhouette score plotted in 
Fig. \ref{fig:silhouette} for fixing the number of clusters.
Observe that the silhouette score is maximized for 3 clusters with a value of 0.872. However, we select the number of clusters to be 7 as maximization of silhouette score is not the only goal. We also need to quantify how many clusters will make the problem tractable {by explaining the DN sufficient}. Note there is a sharp decline in silhouette score if the number of clusters is increased beyond 7.

\begin{figure*}[!htbp]
	\center
	\includegraphics[width=1.07\linewidth]{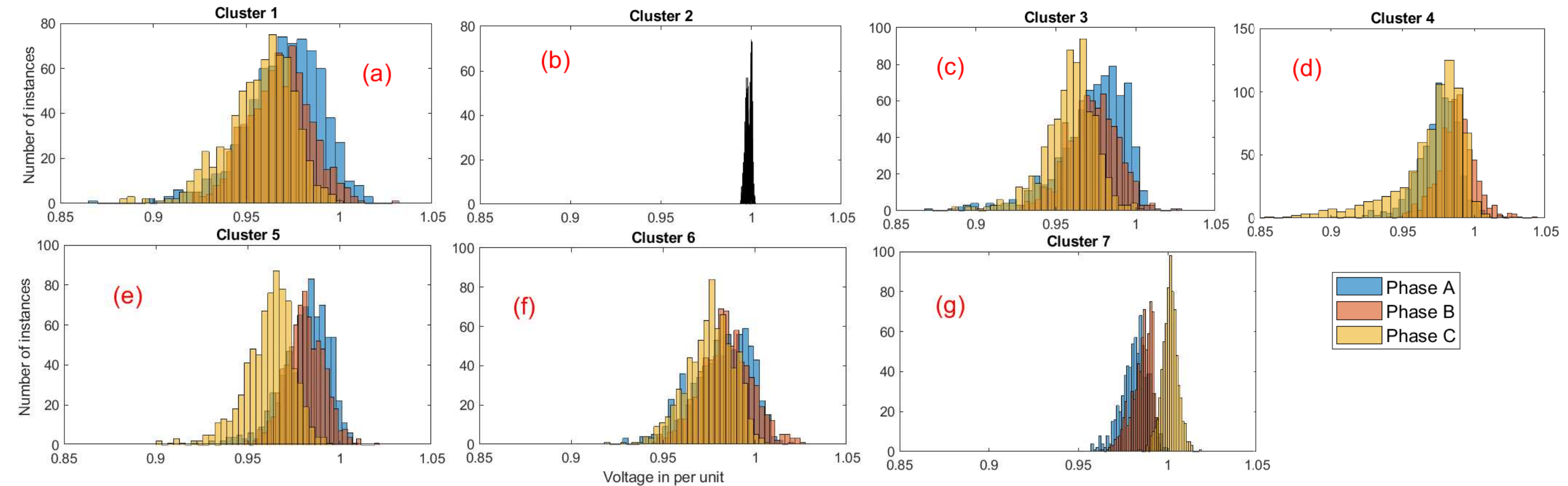}
	\vspace{-2pt}
	\caption{\small{Distribution of voltage in phases in different clusters for Mitnetz Strom DN for cluster 1 to 7.}}
	\label{fig:clustervoltage}
\end{figure*}

Previously, we defined $\beta_c$ as the voltage change threshold for cluster $c$. Note $\beta_c$ will vary with different clusters, as voltage fluctuation in different zones will vary drastically. Fig. \ref{fig:clustervoltage} shows the variation of nodal voltages in seven clusters of the Mitnetz Strom {example LV} distribution network. It can be observed that the voltage variation in cluster 2 is very small, ranging from 0.995 to 1.002. This narrowband is due to cluster 2 including the substation and the slack bus, where the voltage is regulated at 1 per unit level.

\begin{figure}[!htbp]
	\center
	\includegraphics[width=0.8\linewidth]{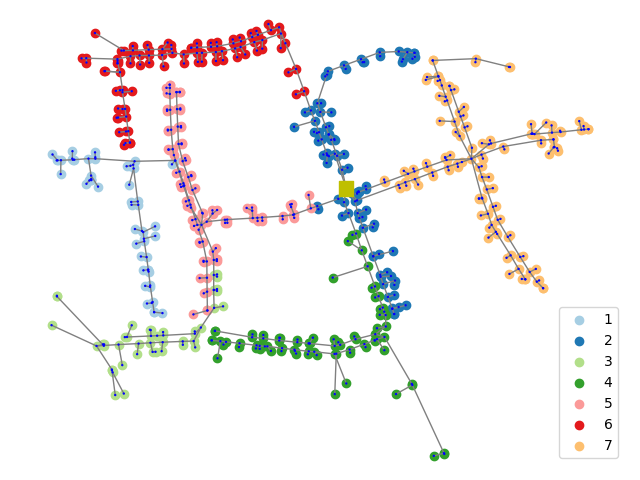}
	\vspace{-10pt}
	\caption{\small{Clusters of DN.}}
	\label{fig:clusterss}
\end{figure}

Fig. \ref{fig:clusterss} shows the DN clusters. We can also comment that the clusters identified are indeed stable, which is validated by 100 Monte Carlo (MC) simulations.
The stability of clusters, impacted due to initializations are discussed in \cite{bubeck2009initialization, kuncheva2006evaluation}.
Clustering based on $k$-means is sensitive to randomized initializations of centroids, and if not stable would provide different clusters in different iterations. 
Observe that the numbering of clusters in Fig. \ref{fig:clusterss} is based on randomized initialization of the centroids, and indeed would vary in a different iteration.

Tab. \ref{tab:phasemapping} details the phase mapping scenarios. In the first case study, we assess the impact of different phase mappings. 
C1 denotes balanced, C2 denotes moderately balanced, and C3 denotes unbalanced phase mapping based on the annual cumulative load on each phase.
For the rest of the paper, if phase mapping is not explicitly mentioned, then C2 phase mapping is used, see Tab. \ref{tab:phasemapping}.

\begin{table}[!htbp]
\centering
\caption{\small{Phase mapping scenarios}}
\begin{tabular}{c|c|c|c|c} 
\cline{3-5}
\multicolumn{1}{l}{} &                   & \multicolumn{3}{l}{Annual cumulative load (MWh)}  \\ 
\hline
ID                   & Cases             & Phase A & Phase B & Phase C                        \\ 
\hline
\hline
C1              & Highly balanced   & 274.6   & 282.3   & 275.7                          \\
\hline
C2              & Fairly balanced   & 288.6   & 292.4   & 251.6                          \\ 
\hline
C3              & Highly unbalanced & 240.4   & 257.7   & 334.4                          \\ 
\hline
\end{tabular}
\label{tab:phasemapping}
\end{table}

The details of the number of consumers, measurement points, and voltage variation for phase mapping C2 (see Table \ref{tab:phasemapping}) are provided in Table \ref{tab:clusterdetails}.

\begin{table}[!htbp]
\centering
\caption{\small{Network attributes {for C2 phase mapping}}}
\begin{tabular}{c|c|c|c|c} 
\hline
\textbf{Cluster ID} & $N_c$ & $M_c$ & $\beta_c$ & {Max voltage deviation}  \\ 
\hline
\hline
1                   & 20                & 4                 & 0.056                          & 0.136                    \\ 
\hline
2                   & 73                & 8                 & 0.008                         & 0.018                   \\ 
\hline
3                   & 21                & 9                 & 0.064                          & 0.164                    \\ 
\hline
4                   & 56                & 2                 & 0.051                         & 0.172                  \\ 
\hline
5                   & 46                & 9                 & 0.0493                      & 0.112                      \\ 
\hline
6                   & 38                & 8                 & 0.040                       & 0.091                     \\ 
\hline
7                   & 59                & 13                & 0.031                          & 0.058                    \\
\hline
\end{tabular}
\label{tab:clusterdetails}
\end{table}

\subsection{Performance of na\"{i}ve model}
\begin{figure}[!htbp]
	\center
	\includegraphics[width=0.5\linewidth]{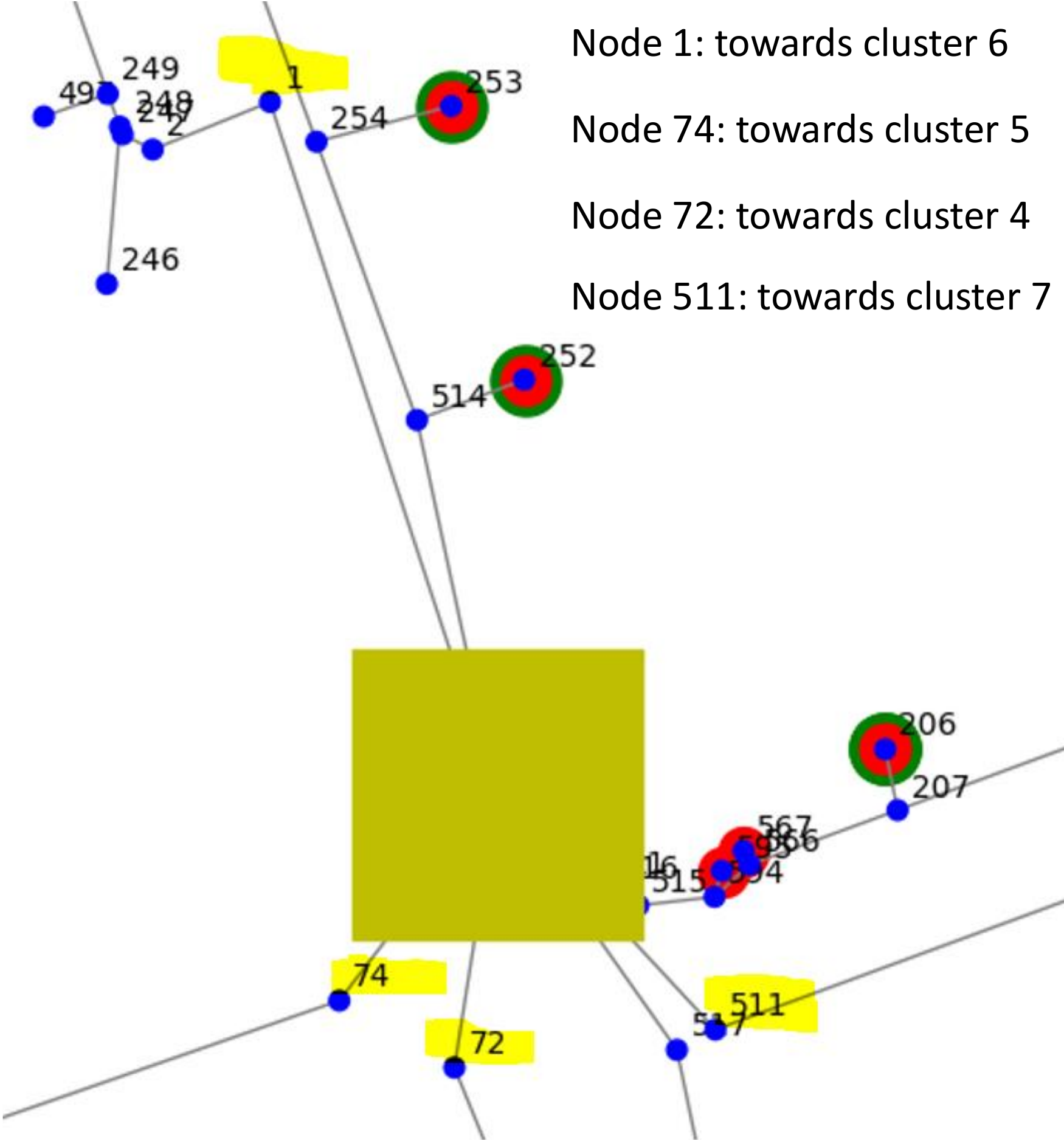}
	\vspace{-10pt}
	\caption{\small{Measurement locations for na\"{i}ve model evaluation}}
	\label{fig:naiveLoc}
\end{figure}

As the majority of prior works utilize a single voltage reference for PCI, we would show the performance of this model, {referred to as the na\"{i}ve model}, prior to evaluation of the proposed phase identification models.
We utilize 4 different measurement points close to the transformer for evaluating the na\"{i}ve phase identification model.
The measurement points are shown in Fig. \ref{fig:measurmentpts}.
It is also indicated that nodes 1, 72, 74, and 511 are connected to feeders going towards clusters 6, 4, 5, and 7 respectively, see Fig. \ref{fig:clusterss}.
A zoomed-in plot of measurement points and the location of the nodes of measurement is shown in Fig. \ref{fig:naiveLoc}.

\begin{figure*}[t]
     \centering
    \begin{subfigure}{0.49\textwidth}
        \raisebox{-\height}{\includegraphics[width=\textwidth]{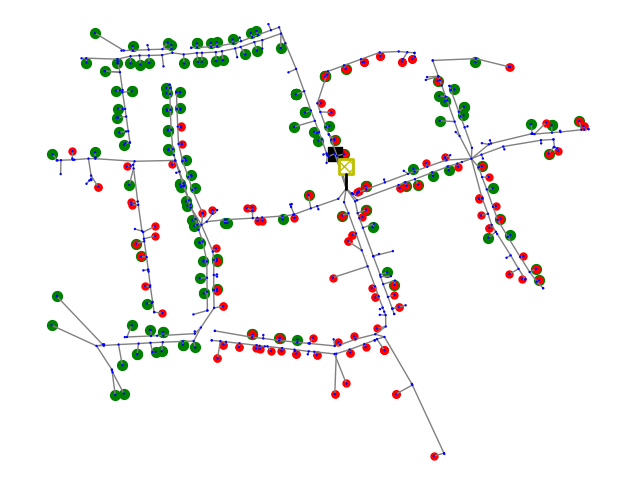}}
        \caption{Node 1 as reference}
    \end{subfigure}
    \hfill
    \begin{subfigure}{0.49\textwidth}
        \raisebox{-\height}{\includegraphics[width=\textwidth]{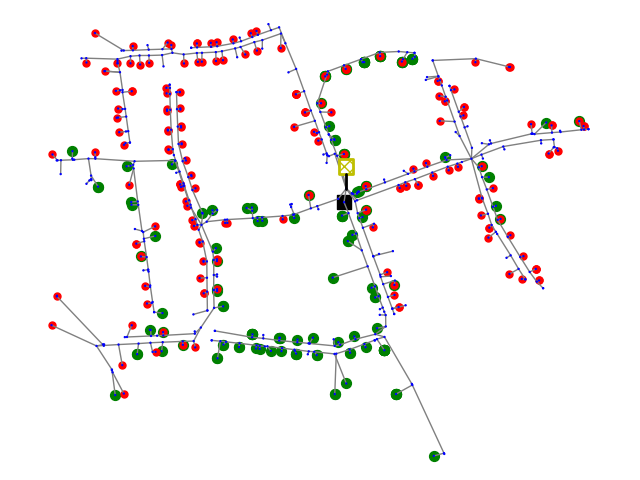}}
        \caption{Node 72 as reference}
    \end{subfigure}
    \begin{subfigure}{0.49\textwidth}
        \raisebox{-\height}{\includegraphics[width=\textwidth]{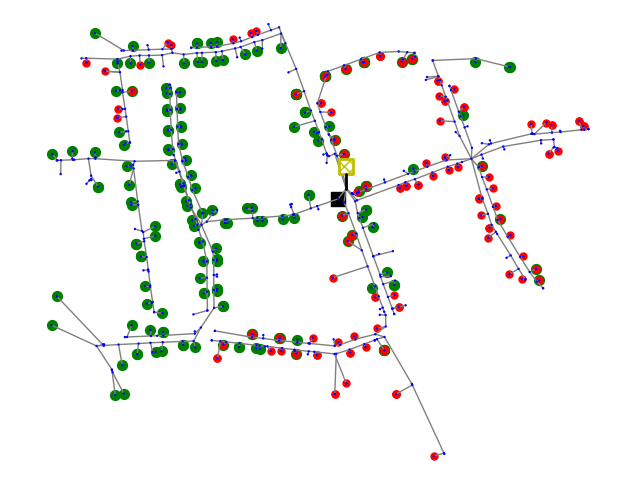}}
    \caption{Node 74 as reference} 
    \end{subfigure}
    \hfill
    \begin{subfigure}{0.49\textwidth}
        \raisebox{-\height}{\includegraphics[width=\textwidth]{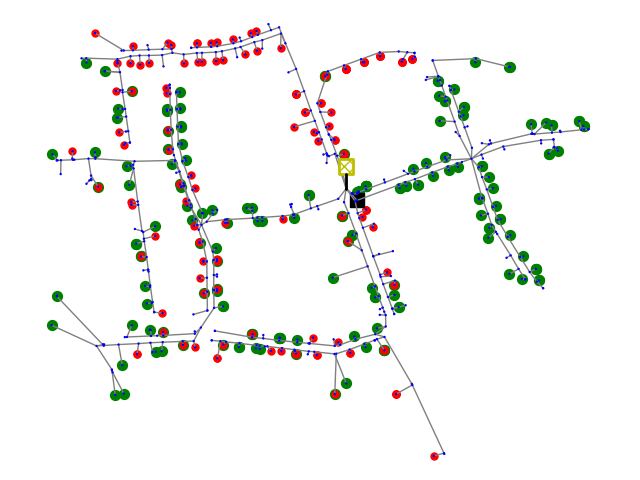}}
    \caption{Node 511 as reference} 
    \end{subfigure}
    \caption{Phase connectivity identification accuracy with na\"{i}ve model. The correct phase estimations are indicated with green circles, and incorrect with red circles. The location of the reference is indicated with a black square. The substation is marked with a yellow square.}
    \label{fig:naiveAccuracy}
\end{figure*}

The results of PCI using the na\"{i}ve model is detailed in Table \ref{tab:naiveres}. 
It lists the cluster-wise PCI accuracy. Observe that na\"{i}ve model correctly estimates the phase connectivity for the cluster with which it is directly connected, however, the estimation accuracy for other clusters can be as low as 0\%.
This is also shown in Fig. \ref{fig:naiveAccuracy}. In Fig. \ref{fig:naiveAccuracy}, the measurement points are indicated by a black square, the correctly estimated consumer phase is indicated by a green circle and red circles show as the incorrect identification.
We can observe that \textit{\textbf{the selection of a reference highly affects the phase connectivity identification accuracy in a multi-feeder distribution network}}.

\begin{table}[!htbp]
\centering
\caption{PCI accuracy with na\"{i}ve model}
\begin{tabular}{c|c|c|c|c} 
\cline{2-5}
\multicolumn{1}{c|}{} & Node 1 & Node 72 & Node 74 & Node 511  \\ 
\hline
\hline
Cluster 1             & 33.33  & 52.94   & \textbf{100 }    & 61.11     \\ 
\hline
Cluster 2             & 44.44  & 50      & 56.76   & 29.03     \\ 
\hline
Cluster 3             & 82.35  & 47.37   & \textbf{100}     & 73.68     \\ 
\hline
Cluster 4             & 12.5   & \textbf{100}     & 35.14   & 57.14     \\ 
\hline
Cluster 5             & 69.77  & 23.26   & \textbf{100}     & 61.36     \\ 
\hline
Cluster 6             & \textbf{100}    & 0       & 63.64   & 15.63     \\ 
\hline
Cluster 7             & 47.06  & 15.56   & 19.57   & \textbf{100 }      \\ 
\hline
overall accuracy      & 47.92  & 44.09   & 55.59 & 52.72     \\
\hline
\end{tabular}
\label{tab:naiveres}
\end{table}

It is clear from the numerical evaluations that
\begin{itemize}
    \item Na\"{i}ve model for phase connectivity identification is very sensitive towards the selection of the reference voltage node which is utilized for phase identification of a single phase consumer, and
    \item Na\"{i}ve model does not consider multiple reference measurements in a DN,
    \item The mean PCI accuracy with na\"{i}ve model in a multi-feeder DN considered in this work is below 56\%.
\end{itemize}
The numerical case studies are presented subsequently.

\subsection{Case study 1: Comparing phase identification models}
Previously, we proposed three metrics $\{J1, J2,J3\}$ and four consensus algorithms $\{S1,S2,S3,S4\}$. In this case study, we compare 12 phase connectivity identification algorithms are proposed and applied to the German DN.
This case study provides recommendations for the best-suited metric and consensus algorithm to be used for phase identification. Note that these recommendations could vary for other DNs. 
All results consider a measurement error of 1\%. In order to eliminate the impact of measurement error on PCI, we perform 1000 MC simulations with different measurement errors calculated using \eqref{eq:measurmenterror}.

\begin{table}[!htbp]
\centering
\caption{Majority rule (S1) model for cluster 2}
\begin{tabular}{c|c|c|c|c} 
\hline
Case                & Metric (Jx) & Accuracy & Confidence factor & Sensitivity  \\ 
           &   & ($\bar{A}^{Jx, Sy}_{c}$) in \% & ($\bar{F}^{ Jx, S_y}_{c}$) & ($D^{ Jx, S_y}_{c}$)  \\ 
\hline
\multirow{3}{*}{C1} & J1           & 50.99  & 0                 & 0.3714       \\ 
\cline{2-5}
                    & J2           & 50.99  & 0                 & 0.3714       \\ 
\cline{2-5}
                    & J3           & 59.99  & 0                 & 0.4641       \\ 
\hline
\multirow{3}{*}{C2} & J1           & 60.76  & 0                 & 0.3878       \\ 
\cline{2-5}
                    & J2           & 60.75  & 0                 & 0.3879       \\ 
\cline{2-5}
                    & J3           & 61.14  & 0                 & 0.4612       \\ 
\hline
\multirow{3}{*}{C3} & J1           & 58.91   & 0                 & 0.3791       \\ 
\cline{2-5}
                    & J2           & 59.83   & 0                 & 0.3792       \\ 
\cline{2-5}
                    & J3           & 59.41  & 0                 & 0.4628       \\
\hline
\end{tabular}
\label{tab:cas1eliminateS1}
\end{table}

The majority rule consensus algorithm (S1) outperforms all other models proposed for all clusters except for cluster 2. For all other clusters, the majority rule provides 100\% accurate rules with a confidence factor of 1.
As detailed earlier, cluster 2 is the part of the DN around the substation. The voltage deviation in this cluster is very small.
Table \ref{tab:cas1eliminateS1} shows the performance of the majority rule consensus algorithm for cluster 2. Observe that all three metrics of phase identification are very poor. Thus, we drop the majority rule consensus algorithm in subsequent evaluations.

\begin{figure}[!htbp]
	\center
	\includegraphics[width=0.8\linewidth]{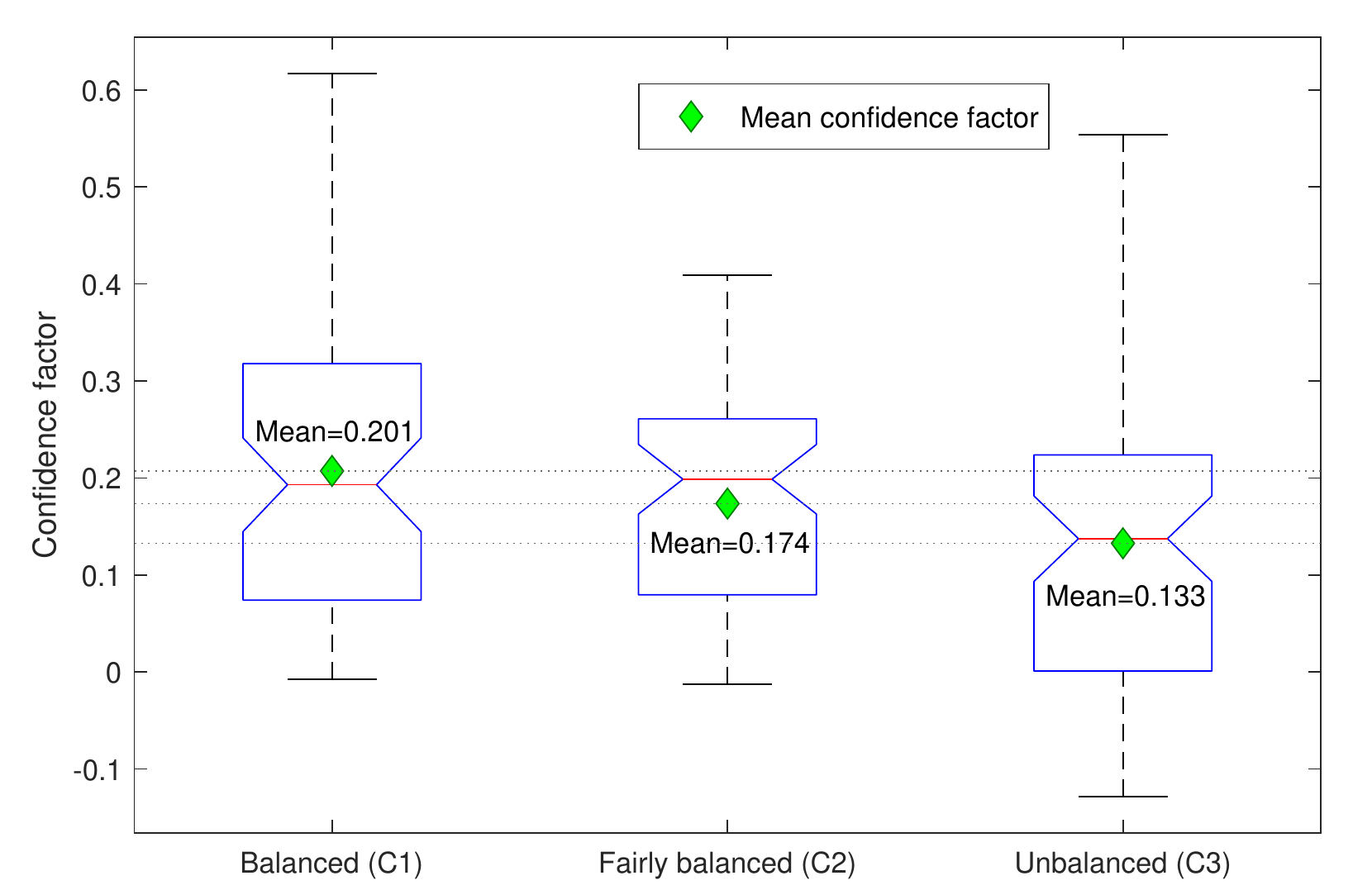}
	\vspace{-2pt}
	\caption{\small{Confidence factor for three-phase mappings for metrics \{J1, J2, J3\} and consensus algorithms \{S2, S3, S4\}.}}
	\label{fig:c1confidence}
\end{figure}

\textcolor{black}{From Fig. \ref{fig:c1confidence}, the confidence factor deteriorates from model C1 which is balanced phase mapping to C3 which is highly unbalanced phase mapping. No noticeable PCI accuracy change is observed for consensus algorithms S2, S3, and S4. 
The mean confidence factor for cases C1, C2, and C3 are 0.201, 0.174, and 0.133 respectively.
Thus, \textbf{\textit{correlation-based phase connectivity identification tends to be more accurate for more balanced phase mapping}}.}

\begin{figure}[!htbp]
	\center
	\includegraphics[width=0.75\linewidth]{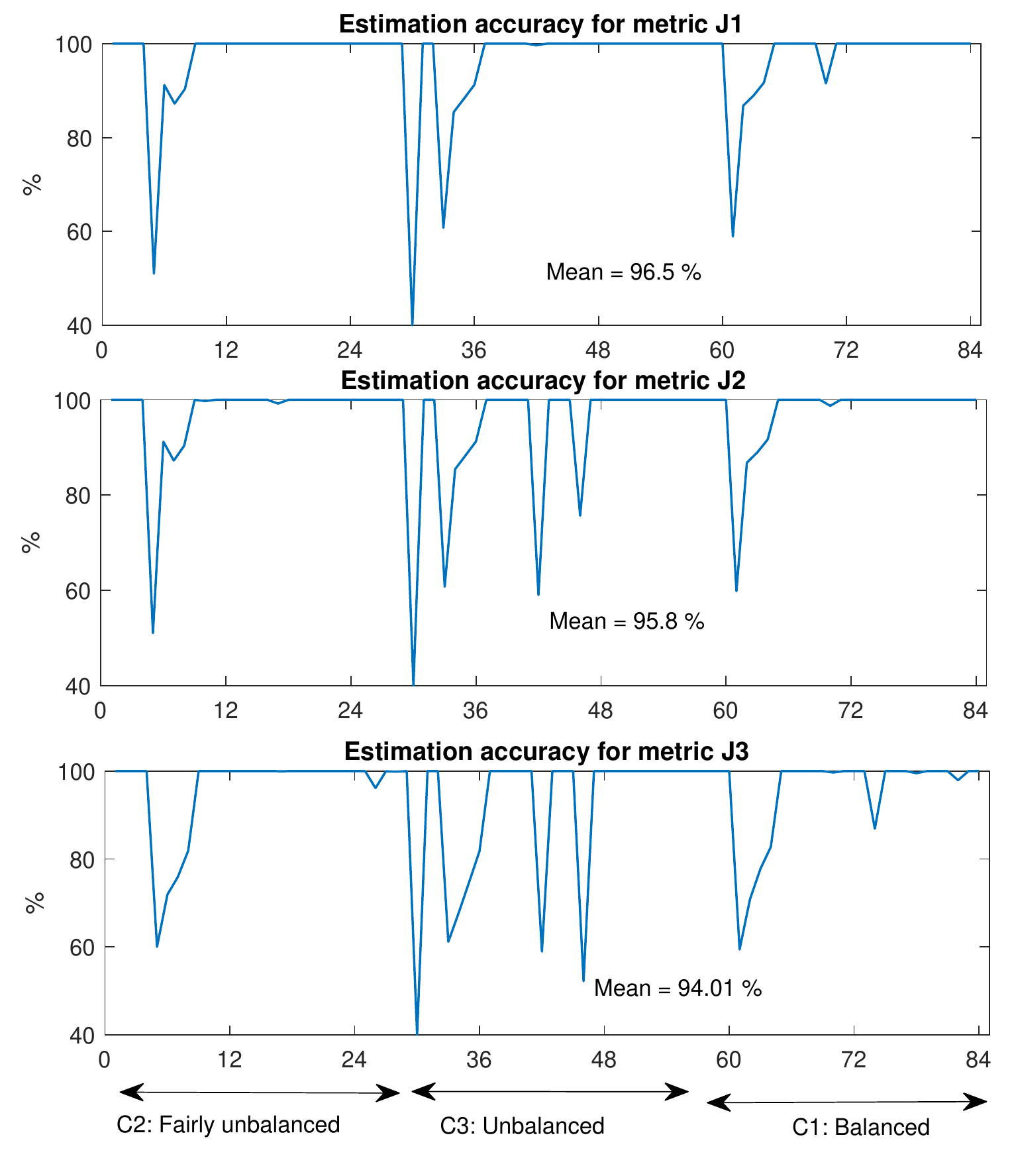}
	\vspace{-5pt}
	\caption{\small{Accuracy for three-phase mappings.}}
	\label{fig:c1accuracy}
\end{figure}

Fig. \ref{fig:c1accuracy} shows the mean phase estimation accuracy for metrics \{J1, J2, J3\}. Observe that mean estimation accuracy is deteriorating for metrics J1 to J3. Further, the accuracy is higher for balanced phase mapping compared to unbalanced phase mapping.
Thus, Fig. \ref{fig:c1confidence} and \ref{fig:c1accuracy} are used to evaluate the impact of phase mappings \{C1, C2, C3\} on phase identification. Further, we observe that metric J1 outperforms others. J1 is more robust as a measure of phase identification.

\begin{figure}[!htbp]
	\center
	\includegraphics[width=0.7\linewidth]{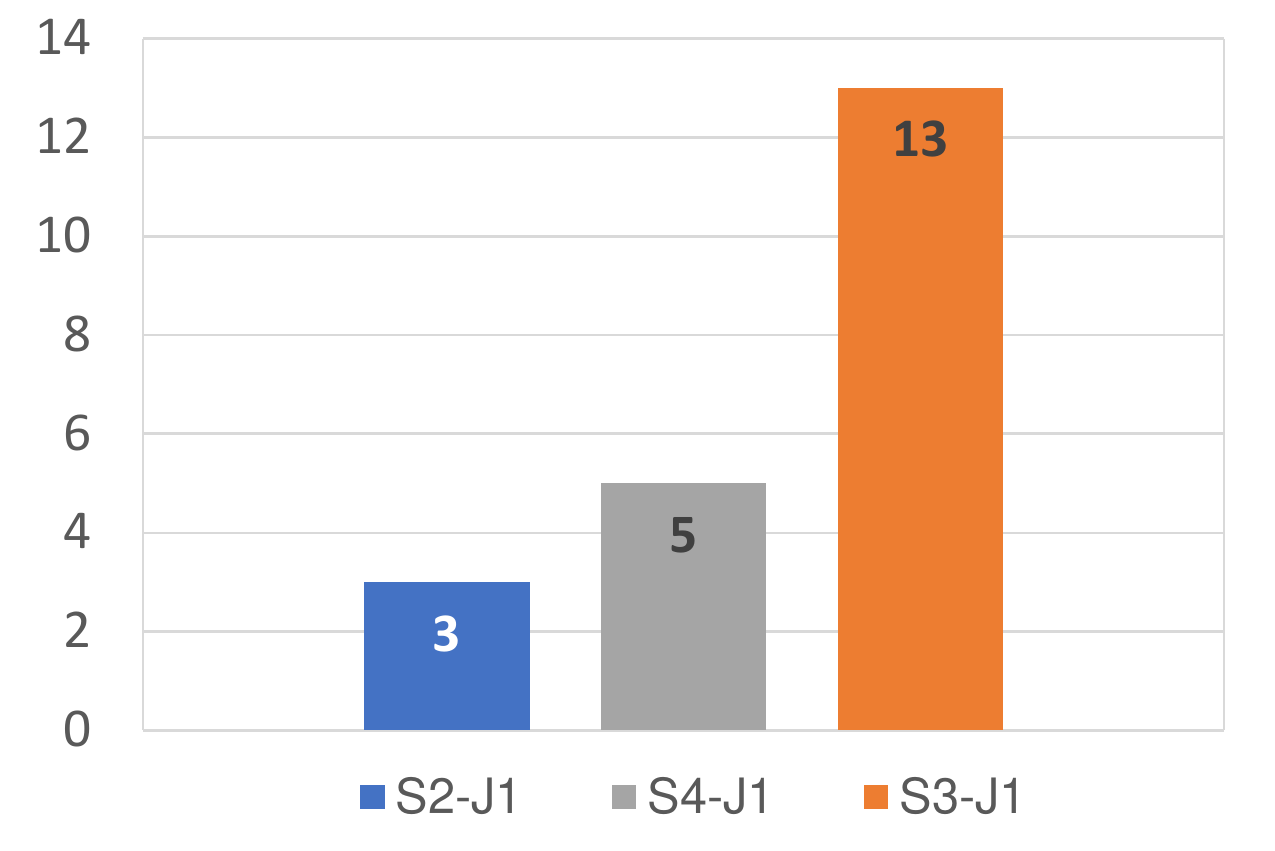}
	\vspace{-2pt}
	\caption{\small{For three different phase mappings and 7 clusters, S3-J1 phase identification model outperforms others for most of the time. The plot shows the histogram  best model for a cluster.}}
	\label{fig:best}
\end{figure}

{
For each cluster \{1,...,7\} and phase mappings \{C1, C2, C3\} we rank the phase identification methods based on the three metrics (a) estimation accuracy in \% denoted as $\bar{A}^{Jx, Sy}_{c}$, (b) confidence factor denoted as $\bar{F}^{ Jx, S_y}_{c}$, and (c) sensitivity towards measurement errors denoted as $D^{ Jx, S_y}_{c}$.}
Out of these 21 models, 13 times consensus algorithm S3 and metric J1 denoted as S3-J1\footnote{Model Sy-Jx denotes that the phase identification algorithm utilizes Jx as the metric and Sy as the consensus algorithm}, outperformed all the other combinations, see Fig. \ref{fig:best}.
Therefore, for all subsequent studies, we will utilize S3-J1 as the best-suited model for phase identification {for the DN considered}.

Table \ref{tab:case1consolidated} shows the consolidated results for metric J1.
Observe that the sensitivity towards phase identification due to measurement error is more than 10 times higher for cluster 2 (close to the substation with low $\beta_c$) compared to other clusters.
This effect can again be attributed to low voltage fluctuations in cluster 2, see Fig. \ref{fig:clustervoltage}.


\begin{table*}[!htbp]
\centering
\footnotesize
\caption{PCI estimation for metric J1}
\begin{tabular}{c|c|c|c|c|c|c|c|c|c|c} 
\cline{3-11}
\multicolumn{1}{c}{} &   & \multicolumn{3}{c|}{Balanced (C1)} & \multicolumn{3}{c|}{Fairly balanced (C2)} & \multicolumn{3}{c|}{Unbalanced (C3)}  \\ 
\hline
Cluster              & Consensus & $\bar{A}^{Jx, Sy}_{c}$      & $\bar{F}^{ Jx, S_y}_{c}$      & $D^{ Jx, S_y}_{c}$        & $\bar{A}^{Jx, Sy}_{c}$      & $\bar{F}^{ Jx, S_y}_{c}$      & $D^{ Jx, S_y}_{c}$               & $\bar{A}^{Jx, Sy}_{c}$      & $\bar{F}^{ Jx, S_y}_{c}$      & $D^{ Jx, S_y}_{c}$           \\ 
\hline
1                    & S2 & 100     & 0.5592  & 0.0295    & 100     & 0.2951  & 0.0344           & 40      & -0.0669 & 0.0386       \\ 
1                    & S3 & 100     & 0.4509  & 0.016     & 100     & 0.376   & 0.0137           & 100     & 0.0615  & 0.0068       \\ 
1                    & S4 & 100     & 0.4361  & 0.0179    & 100     & 0.3723  & 0.017            & 100     & 0.0654  & 0.0076       \\ 
\hline
2                    & S2 & 86.7973 & -0.0022 & 3.6282    & 91.2137 & -0.0027 & 6.94             & 85.4356 & -0.0008 & 4.0251       \\ 
2                    & S3 & 88.9466 & -0.0049 & 0.1293    & 87.2301 & -0.0038 & 0.1338           & 88.3151 & -0.0049 & 0.1344       \\ 
2                    & S4 & 91.6849 & -0.0076 & 0.1202    & 90.3699 & -0.0065 & 0.1249           & 91.2589 & -0.007  & 0.1217       \\ 
\hline
3                    & S2 & 100     & 0.4791  & 0.0264    & 100     & 0.1388  & 0.0567           & 100     & 0.4958  & 0.0261       \\ 
3                    & S3 & 100     & 0.4115  & 0.0144    & 100     & 0.2644  & 0.0103           & 100     & 0.3907  & 0.0138       \\ 
3                    & S4 & 100     & 0.3856  & 0.0159    & 100     & 0.2569  & 0.0122           & 100     & 0.3575  & 0.0157       \\ 
\hline
4                    & S2 & 91.5196 & -0.0035 & 7.2585    & 100     & 0.1442  & 0.069            & 99.6857 & 0.0337  & 3.4869       \\ 
4                    & S3 & 100     & 0.1738  & 0.0166    & 100     & 0.2467  & 0.0225           & 100     & 0.1704  & 0.0168       \\ 
4                    & S4 & 100     & 0.1722  & 0.01471   & 100     & 0.261   & 0.236            & 100     & 0.1607  & 0.0172       \\ 
\hline
5                    & S2 & 100     & 0.3729  & 0.0399    & 100     & 0.2052  & 0.0558           & 100     & 0.2329  & 0.0422       \\ 
5                    & S3 & 100     & 0.3827  & 0.0235    & 100     & 0.3451  & 0.02             & 100     & 0.2453  & 0.0159       \\ 
5                    & S4 & 100     & 0.3783  & 0.0265    & 100     & 0.3287  & 0.0242           & 100     & 0.2245  & 0.0194       \\ 
\hline
6                    & S2 & 100     & 0.2579  & 0.0509    & 100     & 0.2887  & 0.0466           & 100     & 0.1096  & 0.1027       \\ 
6                    & S3 & 100     & 0.3457  & 0.0191    & 100     & 0.2927  & 0.0201           & 100     & 0.2376  & 0.0196       \\ 
6                    & S4 & 100     & 0.3153  & 0.0235    & 100     & 0.2923  & 0.0226           & 100     & 0.1968  & 0.0217       \\ 
\hline
7                    & S2 & 100     & 0.0741  & 0.0961    & 99.9983 & 0.0746  & 0.5477           & 100     & 0.0406  & 0.1595       \\ 
7                    & S3 & 100     & 0.1803  & 0.0461    & 100     & 0.1     & 0.0347           & 100     & 0.2281  & 0.0458       \\ 
7                    & S4 & 100     & 0.1732  & 0.0491    & 100     & 0.0955  & 0.0479           & 100     & 0.1587  & 0.0484       \\
\hline
\end{tabular}
\label{tab:case1consolidated}
\end{table*}

\subsection{Case study 2: Effect of the proximity of measurement on PCI}
\begin{figure}[!htbp]
	\center
	\includegraphics[width=0.7\linewidth]{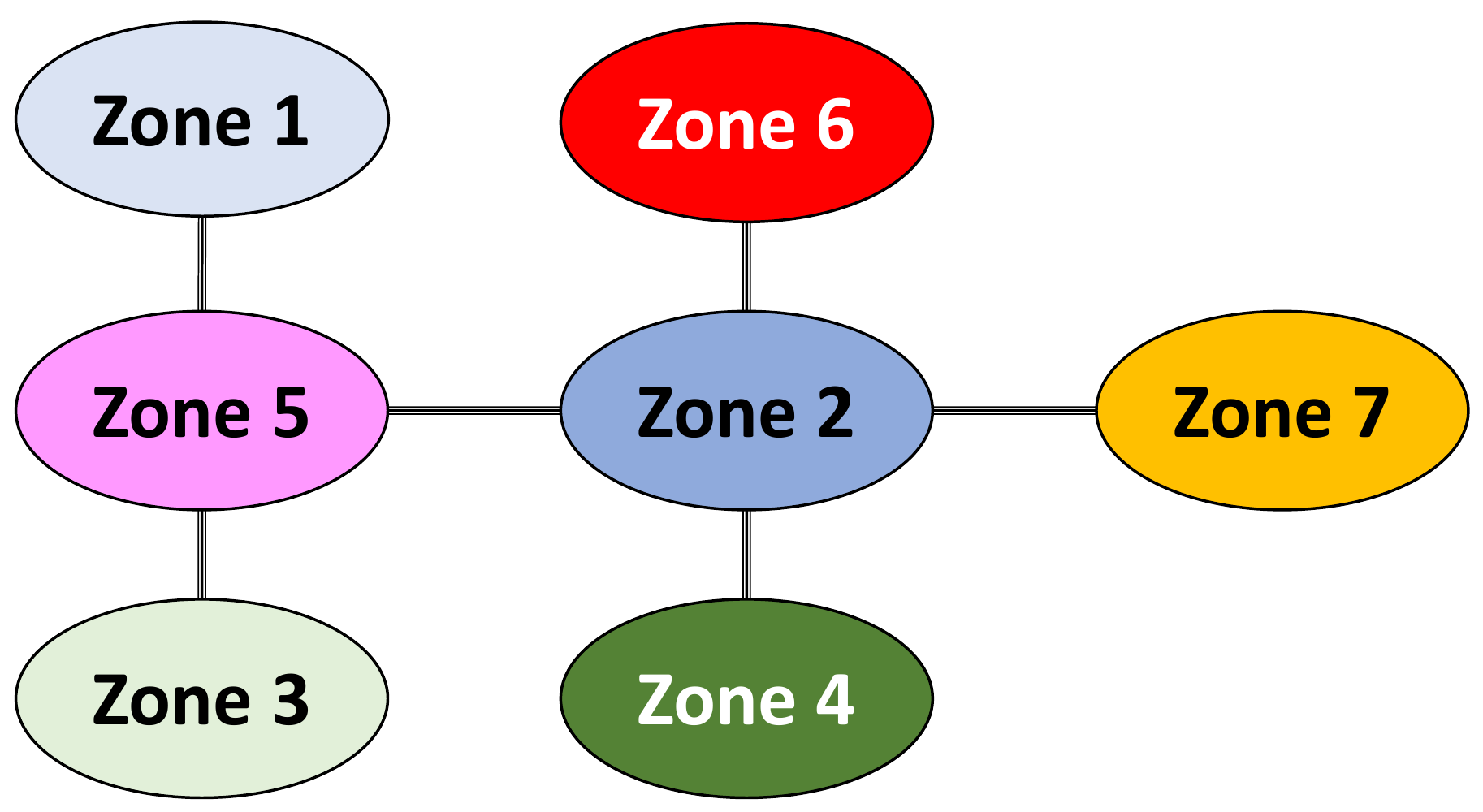}
	\vspace{-2pt}
	\caption{\small{Zones in a simplified network diagram.}}
	\label{fig:zonessimplified}
\end{figure}

The goal of this case study is to assess the impact of measurement point proximity on phase identification {performance metrics}. 
The simplified representation of the original network model in Fig. \ref{fig:clusterss} is shown in Fig. \ref{fig:zonessimplified}.

\begin{table}[!htbp]
\centering
\caption{\small{Zonal connections}}
\begin{tabular}{c|c|c|c|c|c|c|c} 
\hline
L0  & 1         & 2           & 3         & 4         & 5         & 6         & {7}  \\ 
\hline
\hline
L1              & 5         & {[}4,5,6,7] & 5         & 2         & {[}1,2,3] & 2         & 2           \\ 
\hline
L2              & {[}2,3]   & {[}1,3]     & {[}1,2]   & {[}5,6,7] & {[}4,6,7] & {[}4,5,7] & {[}4,5,6]   \\ 
\hline
L3              & {[}4,6,7] & -           & {[}4,6,7] & {[}1,3]   & -         & {[}1,3]   & {[}1,3]     \\
\hline
\end{tabular}
\label{tab:zoneorder}
\end{table}

Zonal connections are listed in Table \ref{tab:zoneorder}. L0 denotes the zone where the $1-\phi$ consumer is located. L1 denotes the zones which are directly connected to L0. Similarly, L2 denotes zones that are not connected directly to L0 but via L1; second-order neighbor, and so on.

\begin{table}[!htbp]
\centering
\caption{Mean PCI accuracy with reference selection}
\begin{tabular}{c|c|c|c|c|c|c|c} 
\hline
level & 1     & 2     & 3     & 4     & 5     & 6     & 7      \\ 
\hline
\hline
L0         & 100   & 90.44 & 100   & 100   & 100   & 100   & 100    \\ 
\hline
L1         & 100   & 43.94  & 100   & 100 & 100 & 99.39 & 78.78  \\ 
\hline
L2         & 100 & 35.71   & 100 & 29.57 & 44.94 & 75.18 & 48.44  \\ 
\hline
L3         & 34.76 & -     & 47.31 & 27.37 & -     & 30.47 & 19.55 \\
\hline
\end{tabular}
\label{tab:c2r1}
\end{table}

\begin{table}[!htbp]
\centering
\caption{Mean confidence factor ($\bar{F}^{ Jx, S_y}_{c}$) with reference selection}
\begin{tabular}{c|c|c|c|c|c|c|c} 
\hline
level & 1      & 2      & 3      & 4      & 5      & 6      & 7       \\ 
\hline
\hline
L0         & 0.372  & -0.007 & 0.256  & 0.261  & 0.329  & 0.293  & 0.096   \\ 
\hline
L1         & 0.357  & -0.005 & 0.262  & 0.141  & 0.317  & 0.039  & -0.003  \\ 
\hline
L2         & 0.333  & -0.005 & 0.240  & -0.004 & -0.014 & -0.025 & -0.005  \\ 
\hline
L3         & -0.049 & -      & -0.051 & -0.004 & -      & -0.016 & -0.003  \\
\hline
\end{tabular}
\label{tab:c2r2}
\end{table}

\begin{table}[!htbp]
\centering
\caption{Mean PCI STD ($D^{ Jx, S_y}_{c}$) with reference selection}
\begin{tabular}{c|c|c|c|c|c|c|c} 
\hline
level & 1     & 2     & 3     & 4     & 5     & 6     & 7      \\ 
\hline
\hline
L0         & 0.017 & 0.125 & 0.012 & 0.023 & 0.024 & 0.022 & 0.048  \\ 
\hline
L1         & 0.019 & 0.223 & 0.018 & 0.074 & 0.039 & 0.078 & 0.108  \\ 
\hline
L2         & 0.044 & 0.259 & 0.042 & 0.099 & 0.088 & 0.100 & 0.124  \\ 
\hline
L3         & 0.068 & -     & 0.066 & 0.094 & -     & 0.101 & 0.116  \\
\hline
\end{tabular}
\label{tab:c2r3}
\end{table}

Tables \ref{tab:c2r1}, \ref{tab:c2r2}, and \ref{tab:c2r3} list the numerical results with mean phase estimation metrics for each consumer and each measurement point in the DN. 
The following observations are made:
\begin{itemize}
    \item Phase estimation accuracy deteriorates as the electrical distance between the measurement point used as the reference in correlation-based phase identification and the $1-\phi$ consumer with unknown phase connection increases,
    \item The confidence of estimation deteriorates as the distance between measurement and consumer increases,
    \item The mean estimation variance increases (implying estimation becomes more sensitive to measurement errors)  as the distance between the measurement point used as the reference in correlation based phase identification and the single phase consumer with unknown phase connection increases.
\end{itemize}

The phase connectivity identification metrics proposed in this work not only quantitatively provide the estimation accuracy in \% but also qualitatively provide the confidence factor (the higher the better) and the sensitivity to measurement errors (the lower the better). Using these metrics, we observe that selecting of measurement points in proximity is better in terms of phase estimation quality. This also \textbf{validates our zonal phase identification approach} we have established in this work.

\subsection{Case study 3: Impact of measurement error}

\begin{figure}[!htbp]
	\center
	\includegraphics[width=0.75\linewidth]{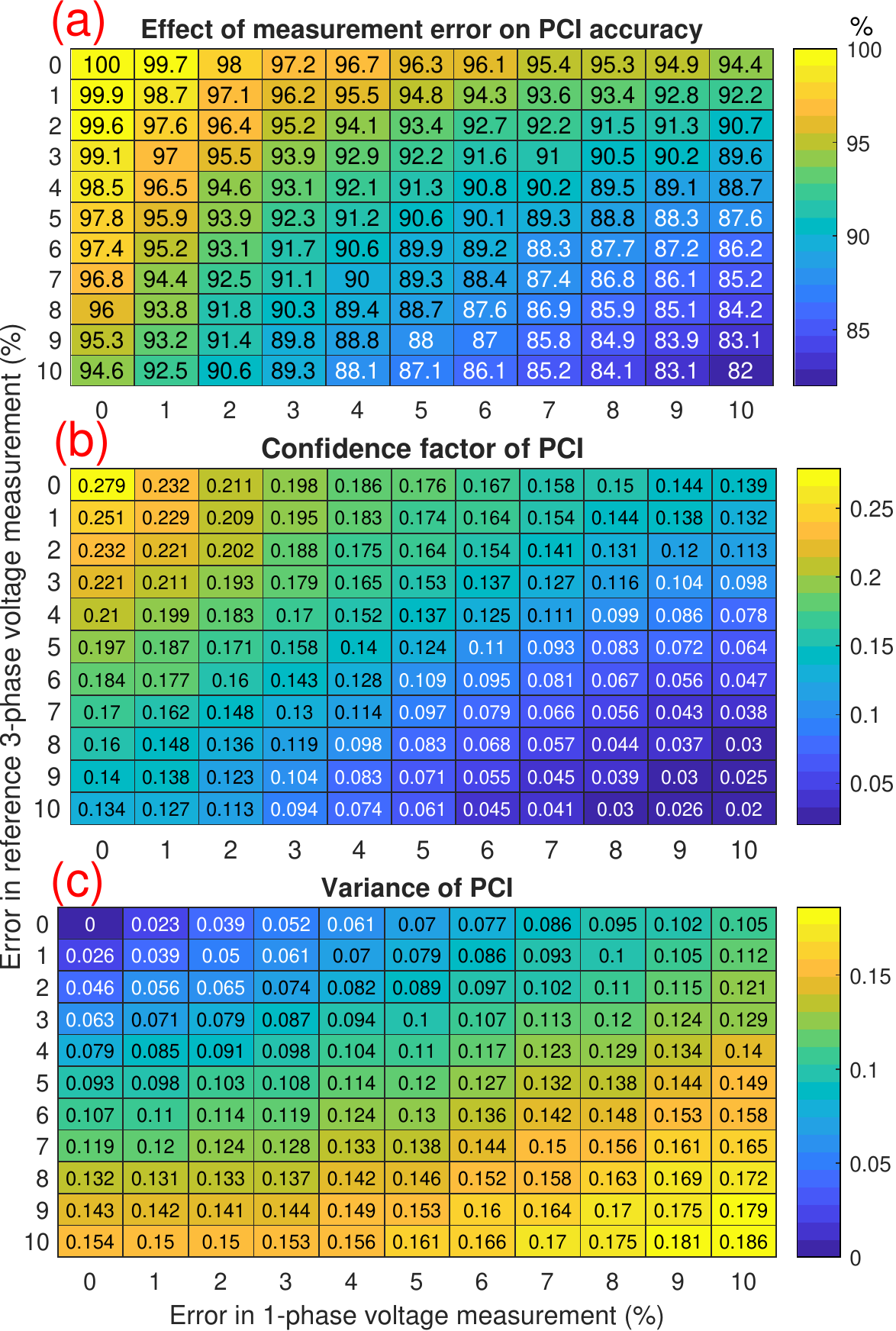}
	\vspace{-2pt}
	\caption{\small{Impact of measurement accuracy on PCI metrics. (a) shows the accuracy, (b) shows the confidence factor and (c) shows the variance of PCI respectively.}}
	\label{fig:case31}
\end{figure}

In case study 1, we identified the best-suited model for phase identification as the S3-J1. Previously, we observed that an imbalance in DN leads to worse estimation compared to balanced phase mapping. In order to not have pessimistic or optimistic phase identification results, we utilize C2 phase mapping. We apply different levels of measurement errors at the point of reference and at the point of single phase {consumer point of} connection.
In order to not be biased by one sample of estimation error, we perform 1000 MC simulations.
The three-phase identification metrics are shown in Fig. \ref{fig:case31}.
The three metrics are almost equally influenced by measurement error at reference voltage measurement or 1-$\phi$ consumer end. 
Observe from Fig. \ref{fig:case31}
\begin{itemize}
    \item Mean phase estimation accuracy for 1\% error in reference voltage and consumer voltage measurement is 98.7\%. Implying, our consensus-based phase estimation algorithm was able to estimate more than 308 out of 313 of single-phase consumer phases accurately on average.
    \item For 10\% error in reference voltage and consumer voltage measurement is still 82\%. Thus, the phase estimation proposed in this work is largely immune to measurement errors.
    \item The confidence factor deteriorates with an increase in measurement errors, see Fig. \ref{fig:case31}(b).
    \item The sensitivity of phase estimation towards measurement error, shown as the variance in estimation Fig. \ref{fig:case31}(c) increases with an increase in measurement error.
\end{itemize}

\subsection{Case study 4: Effect of size of neutral conductor}
\begin{figure}[!htbp]
	\center
	\includegraphics[width=0.5\linewidth]{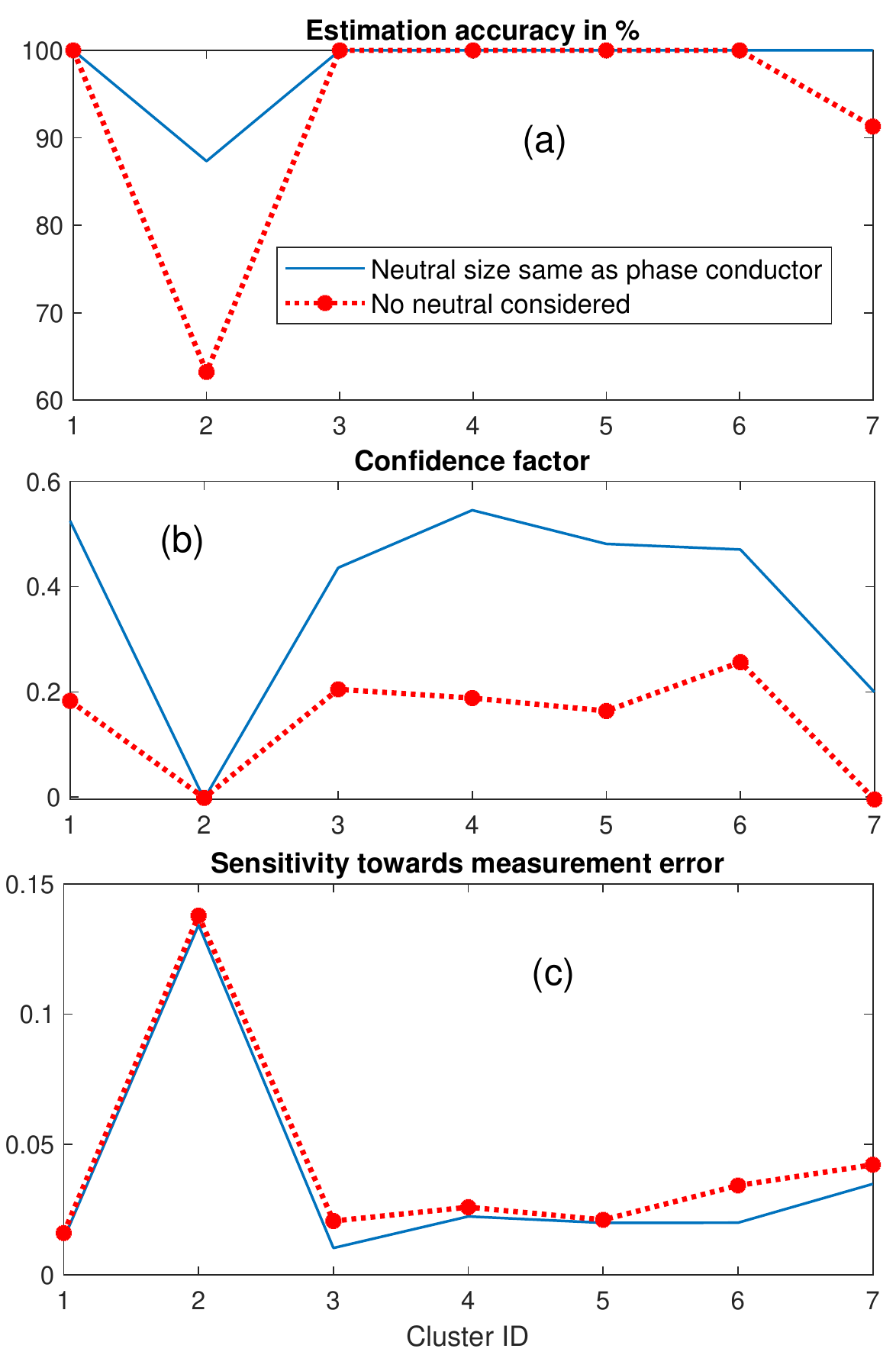}
	\vspace{-2pt}
	\caption{\small{Impact of neutral conductor modeling on phase estimation metrics.}}
	\label{fig:case41}
\end{figure}

European DNs are 4-wire systems with a neutral conductor. The impact of modeling the neutral conductor is not assessed in any of the prior works. This is especially crucial if the digital twin is used for generating synthetic data.
Fig. \ref{fig:case41} shows the phase identification metrics for model S3-J1 and phase mapping C2 with and without neutral conductor considerations.
It is clear that modeling the neutral conductor improves estimation accuracy for all three metrics.

\pagebreak

\section{Conclusions}
\label{section7}

We propose a phase connectivity identification (PCI) algorithm that utilize voltage time series, distribution network (DN) zones, and multiple measurements as references for improving the PCI accuracy. 
The proposed phase identification algorithm builds a consensus among multiple estimations in a zone. This method is extended to consider metrics derived from voltage time series, which filters larger voltage deviations.
Due to the consideration of multiple measurements, the PCI is drastically exceeding the performance compared to the widely used na\"{i}ve model. Further, our approach is immune to network topology and measurement errors, as PCI accuracy is not dependent on only one reference measurement point.
We also propose phase connectivity identification metrics that not only quantitatively describe the estimation accuracy, but also qualitatively describe how good the PCI is.
We utilize a real German DN with limited observability for phase identification. This network has 602 nodes and 313 single-phase consumers. 
It is observed that the original voltage time series without salient feature extraction and absolute weighted consensus algorithm outperforms all the other phase estimation algorithms.
The proposed algorithm identifies the phase connections on average of over 308 consumers accurately for 1\% measurement error at the consumer end and reference measurements for 1000 Monte Carlo simulations. Thus, the proposed algorithm is robust towards uncertainty with high precision.

{In future work, we will consider synchronization errors in measurement while limiting DN observability even further. Further assessment is needed for selecting the best-suited algorithm for phase identification with varying network layouts, load profiles, and PV penetration. Finally, we will extend this work for executing the algorithms on real measurement data and verify algorithm efficacy by intrusive phase measurements.}

\section*{Acknowledgement}
This work is supported by the H2020 EUniversal project, grant agreement ID: 864334 (\url{https://euniversal.eu/}).
We would like to thank Clara Gouveia and Gil Silva Sampaio at INESC TEC, Porto for their comments on problem formulation.
We would like to thank Marta Vanin (KU Leuven), Deepjyoti Deka (Los Alamos National Lab), Lucas Pereira (Técnico Lisboa) for their insightful comments on the paper.
Special thanks to Kseniia Sinitsyna at Mitnetz Strom for her help in data handling.

\pagebreak

\bibliographystyle{elsarticle-num}
\bibliography{reference}



\end{document}